# Normal-state nodal electronic structure in underdoped high-$T_\mathrm{c}$ copper oxides

Suchitra E. Sebastian,[1] N. Harrison,[2] F. F. Balakirev,[2] M. M. Altarawneh,[2,3]
P. A. Goddard,[4] Ruixing Liang,[5,6] D. A. Bonn,[5,6] W. N. Hardy,[5,6] G. G. Lonzarich[1]

[1]Cavendish Laboratory, Cambridge University, JJ Thomson Avenue, Cambridge CB3 OHE, U.K.,
[2]National High Magnetic Field Laboratory, LANL, Los Alamos, NM 87504,
[3]Department of Physics, Mu'tah University, Mu'tah, Karak, 61710, Jordan,
[4]Department of Physics, University of Warwick, Gibbet Hill Road, Coventry, CV4 7AL, U.K.,
[5]Department of Physics and Astronomy, University of British Columbia, Vancouver V6T 1Z4, Canada,
[6]Canadian Institute for Advanced Research, Toronto M5G 1Z8, Canada,

**An outstanding problem in the field of high-transition-temperature (high $T_\mathrm{c}$) superconductivity is the identification of the normal state out of which superconductivity emerges in the mysterious underdoped regime.[1] The normal state uncomplicated by thermal fluctuations is effectively accessed by the use of applied magnetic fields sufficiently strong to suppress long-range superconductivity at low temperatures.[2,3] Proposals in which the normal ground state is characterised by small Fermi surface pockets that exist in the absence of symmetry breaking[1,4–8] have been superseded by models based on the existence of a superlattice that breaks the translational symmetry of the underlying lattice.[7–15] Recently, a charge superlattice model that positions a small electron-like Fermi pocket in the vicinity of the nodes (where the superconducting gap is minimum)[8,9,16,17] has been proposed a replacement for the prevalent superlattice models[10–14] that position the Fermi pocket in the vicinity of the pseudogap at the antinodes (where the superconducting gap is maximum).[18]**

**Although some ingredients of symmetry breaking have been recently revealed by crystallographic studies, their relevance to the electronic structure remains unresolved.[19–21] Here we report angle-resolved quantum oscillation measurements in the underdoped copper oxide $YBa_2Cu_3O_{6+x}$. These measurements reveal a normal ground state comprising electron-like Fermi surface pockets located in the vicinity of the superconducting gap minima (or nodes), and further point to an underlying superlattice structure of low frequency and long wavelength with features in common with the charge order identified recently by complementary spectroscopic techniques.[14, 19–22]**

The normal ground state electronic structure revealed by our measurements is summarised in Fig. 1. Quantum oscillations measured in the contactless electrical resistivity of the underdoped cuprate $YBa_2Cu_3O_{6.56}$ are shown in Fig. 2**a**. The prominent oscillatory beat structure reveals a sizeable frequency spread.[8, 16] Measurements are made as a function of magnetic field up to 85 T over a wide range of field orientations as defined by the polar angle $\theta$ and the azimuthal angle $\phi$. This data may be used to identify characteristics of the momentum-space electronic structure in the way that, for example, X-ray data may be used to identify the real-space lattice structure of a crystal. In particular, it can be employed to identify the correct Fermi surface model.[23] In the case of underdoped $YBa_2Cu_3O_{6+x}$, which has a primitive orthorhombic lattice, Fig. 2**b** illustrates the quasi-two dimensional electronic structure characterised by a cylindrical Fermi surface with fundamental neck and belly warping that would be expected in the absence of a superlattice (Fig. 3**a-d**). A distinguishing characteristic of this neck and belly Fermi surface is that it gives rise to a resonance in the amplitude of quantum oscillations at a polar angle known as the Yamaji angle.[23] The resonance in amplitude is expected to grow with the size of the frequency spread, and its location is expected at approximately is expected at approximately $60°$ in $YBa_2Cu_3O_{6+x}$ given the measured mean diameter of the cylinder corresponding to the dominant quantum oscillation frequency, and the $c$-axis lattice constant (Fig. 2**b**,**c**). Strikingly, however, despite the sizeable frequency spread of the quantum oscillation spectrum (Fig. 2**a**), higher angle data in Fig. 2**a** reveal the Yamaji resonance anticipated for a fundamental neck and belly Fermi surface geometry to be absent, pointing to an alternative Fermi surface geometry. Supporting analyses are shown in Methods and Extended Data Figs. 1,2,3,4.

To arrive at the correct Fermi surface model we are guided by the following key experimental findings: (1) as shown in Fig. 4**a**, the replacement of the expected Yamaji resonance indeed by an anti-resonance; (2) as shown in Fig. 4**b**, the fourfold anisotropy in $\phi$ dependence of the quantum oscillation amplitude for different values of $\theta$, with a maximisation of quantum oscillation amplitude along the $a$ and $b$ crystallographic directions and minimisation along the diagonal directions, for example, $\phi = 45°$; and (3) as shown in Fig. 4**b**; the enhancement in fourfold anisotropy as a function of azimuthal angle with increasing values of polar angle $\theta$.

While the observed experimental features are in sharp variance with those expected for a dominant fundamental neck and belly Fermi surface geometry, we find that a dominant staggered twofold Fermi surface with warping maxima along the diagonal directions in momentum space (Fig. 1**a-c**, Fig. 3**f-h**) predicts the experimental features that we have observed. The simulation of the quantum oscillation waveform (Fig. 3**h**, and Fig. 4, and Methods) shows that such a staggered twofold Fermi surface geometry yields first an anti-resonance in the quantum oscillation amplitude at a special value of $\theta$ along the diagonal directions in $\phi$, second, a fourfold anisotropy in the quantum oscillation amplitude as a function of $\phi$ that is minimised along the diagonal directions, and finally an enhancement in the fourfold anisotropy as a function of $\phi$ with increased values of $\theta$. Layered materials with examples of both fundamental neck and belly and staggered twofold Fermi surface geometries are discussed in Methods.

Simulations of a staggered twofold Fermi surface geometry with modulation amplitude $\Delta F_{\text{twofold}} \approx 15$ T, agree well with the measured quantum oscillation data as a function of all three parameters $B$, $\theta$, and $\phi$ (Fig. 4, Extended Data Table 2). In addition to the ampli-tude damping factor for a staggered twofold geometry ($R_\text{w}^\text{twofold}$), conventional damping factors are included that are used to describe layered materials, which arise from thermal smearing, impurity scattering or quenched inhomogeneities, Zeeman splitting and magnetic breakdown. The quantum oscillatory frequency spread is captured by Fermi surface splitting from a finite bilayer or spin orbit coupling. More details of the simulation are given in Methods, Extended Data Figs. 2**c**,4,5,6,7, and Extended Data Tables 1,2.

The staggered twofold Fermi surface geometry we observe would not be expected to dominate within the primitive orthorhombic unit cell of underdoped $YBa_2Cu_3O_{6+x}$. Yet just such a geometry would arise from the unique symmetry of the corner T point in the Brillouin zone of a body-centred orthorhombic unit cell defined in Figs. 1 and 3**e-h** (see Methods). At this special corner T point, the twofold in-plane rotational symmetry alternates by 90° between adjacent symmetry planes. The diagonal orientation of the maximal warping directions of the Fermi surface (shown in Figs. 1**a-c** and identified from the azimuthal anisotropy in Fig. 4**b**), reveals the new body-centred orthorhombic Brillouin zone to be oriented concentrically with the original primitive orthorhombic Brillouin zone, and to be defined by orthogonal ordering wavevectors $\mathbf{Q}_1$ and $\mathbf{Q}_2$. The staggered twofold Fermi surface pockets at the corner T point are therefore located in the nodal region of the original Brillouin zone. A nodal Fermi surface pocket contained within the reconstructed Brillouin zone is not only consistent with the present measurements, but also with the low observed value of linear specific heat coefficient in high magnetic fields[6] and strong chemical potential oscillations inferred from previous quantum oscillation studies.[17]

We now consider the possible origin of the superlattice responsible for the above nodal staggered twofold Fermi surface pockets, which emerge from a reconstruction of a large Fermi surface determined from band structure, characteristic of the original primitive orthorhombic lattice in the normal state.[8] Numerous proposals for superlattices in the copper oxides have been put forward, for example in refs.[1,7–16,24–29] Of particular relevance to our observations is the charge order recently detected in $YBa_2Cu_3O_{6+x}$ by techniques such as X-ray diffraction,[19,20] ultrasound,[21] nuclear magnetic resonance,[14] and optical reflectometry.[22] These observations point to a superlattice characterised by the same ordering wave vectors $\mathbf{Q}_1$ and $\mathbf{Q}_2$ identified by our measurements and defined in Figs. 1, with the superlattice scaling parameters $\delta_1 \approx \delta_2 \approx 0.3$, and anisotropic amplitudes in some cases. We note that for Fermi surface reconstruction, the superlattice need not be strictly long-range or static, but it must not be fluctuating over a range much smaller than the cyclotron radius, nor with a frequency much larger than the cyclotron frequency.

Numerical calculations indeed show that this type of charge order can give rise to nodal Fermi surface pockets similar to those observed in our measurements, potentially accompanied by antinodal gaps in the electronic excitation spectrum[8,9,24] (schematic shown in Extended Data Figs. 6, 7). Furthermore, the resulting nodal electron-like Fermi surface pockets yield the observed negative sign of Hall coefficient in the accessed high magnetic field limit (see Methods, refs.[8–10]) in contrast to the positive value expected for the large unreconstructed Fermi surface in the absence of the superlattice. Our measurements do not distinguish between charge order with s-wave symmetry and unconventional symmetry such as d-wave symmetry (i.e., bond order). Further, they are compatible with charge modulation components (of wavevectors $\mathbf{Q}_1$ and $\mathbf{Q}_2$) not only of similar amplitudes, but also of significantly different amplitudes, as in a nearly uniaxial structure.[31]

The strength of magnetic fields in which the present measurements were carried out (up to 85 T in Fig. 2**a** and up to 100 T in Extended Data Fig. 3**b**) is adequate to suppress long-range superconducting order, and thus reveal the normal ground state underpinning superconductivity in underdoped $YBa_2Cu_3O_{6+x}$. The resulting normal ground state is a Fermi liquid characterised by staggered twofold Fermi pockets that emerge naturally from a long-range static or short-range slowly fluctuating superlattice defined by ordering wavevectors $\mathbf{Q}_1$ and $\mathbf{Q}_2$ (Fig. 1), support for which is found from recent complementary experiments.[14,19–21] Superconductivity may therefore be viewed as emerging from the pairing of quasiparticles on Fermi pockets we locate in the nodal region of momentum space. Our finding clarifies observations from complementary experiments, for example Raman spectroscopy,[30] which show that in the underdoped regime, Bogoliubov quasiparticles are confined to momentum space islands around the nodal regions of the Brillouin zone. This is in sharp contrast with the emergence of Bogoliubov quasiparticles from both nodal and antinodal regions of the starting large unreconstructed Fermi surface in the overdoped regime.

## Method Summary

Quantum oscillations in the electrical resistivity were measured using a contactless technique on a high quality de-twinned single crystal of $YBa_2Cu_3O_{6.56}$ over a wide range of polar and azimuthal angles, $\theta$ and $\phi$, in magnetic fields up to $\approx 85$ T and at 1.5 K (Fig. 2**2** and Extended Data Fig. 1). Measurements up to 100 T at $\theta = 0$ are presented in Extended Data Fig. 3.

The model used for the quantum oscillation simulations is discussed in the Methods. Of particular importance here is the Fermi surface geometrical damping or warping factor, taken to be of the form $R_\mathrm{w}^\mathrm{twofold} = J_0 \left[ \frac{2\pi \Delta F_\mathrm{twofold}}{B \cos\theta} \sin 2\phi J_2(k_F(c'/2)\tan\theta) \right]$, which is expected for a staggered twofold Fermi surface geometry in a body-centred weakly-orthorhombic unit cell (Fig. 1**a,b**). Here $k_\mathrm{F} \approx 0.13$ Å$^{-1}$ (corresponding to the dominant frequency $F_0 \approx 530$ T) is the average radius of the Fermi surface pockets in the basal plane, $\Delta F_\mathrm{twofold}$ is a measure of the magnitude of the staggered twofold warping and $J_0$ and $J_2$ are the zeroth and second-order Bessel functions, respectively. The results of the simulation are shown and explained in Fig. 4 and in Extended Data Table 1.

For a Fermi surface with fundamental neck and belly warping geometry in a primitive weakly-orthorhombic unit cell (see Fig. 2**b**), a different warping damping factor $R_\mathrm{w}^\mathrm{neck-belly} =$ $J_0 \left[ \frac{2\pi \Delta F_{\text{neck-belly}}}{B \cos\theta} J_0(k_F c \tan\theta) \right]$ is required. The results of a simulation using $R_{\text{w}}^{\text{neck-belly}}$ with model parameters listed in Extended Data Table 3 are shown in Fig. 2**c** and Extended Data Fig. 2**b** and **c**. The striking difference in the angular-dependence of the quantum oscillation waveform corresponding to $R_{\text{w}}^{\text{neck-belly}}$ and $R_{\text{w}}^{\text{twofold}}$ (illustrated in Fig. 3**d** and **h** respectively) is shown in Extended Data Fig. 2.

# References and Notes

[1] Lee, P. A., Nagaosa, N., Wen, X. G., Doping a Mott insulator: Physics of high-temperature superconductivity. *Rev. Mod. Phys.* **78**, 17-85 (2006).

[2] Grissonnanche, G., Cyr-Choiniere, O., Laliberte, F., Rene de Cotret, S., Juneau-Fecteau, A., Dufour-Beausejour, S., Delage, M. -E., LeBoeuf, D., Chang, J., Ramshaw, B. J., Bonn, D. A., Hardy, W. N., Liang, R., Adachi, S., Hussey, N. E., Vignolle, B., Proust, C., Sutherland, M., Kramer, S., Park, J. -H., Graf, D., Doiron-Leyraud, N., Taillefer, L. Direct measurement of the upper critical field in cuprate superconductors. *Nature Commun.* **5**, 4280 (2014); DOI: 10.1038/ncomms4280.

[3] Wang, Y., Ong, N. P., Xu, Z. A., Kakeshita, T., Uchida, S., Bonn, D. A., Liang, R., Hardy, W. N. High Field Phase Diagram of Cuprates Derived from the Nernst Effect. *Phys. Rev. Lett.* **88**, 257003/1-4 (2002).

[4] Yang, K. -Y., Rice, T. M., and Zhang, F. -C. Phenomenological theory of the pseudogap state *Phys. Rev. B* **73** 174501 (2006).

[5] Anderson, P. W., Lee, P. A., Randeria, M., Rice, T. M., Trivedi, N., and Zhang, F. C. The physics behind high-temperature superconducting cuprates: the 'plain vanilla' version of RVB. *J. Phys.: Condens. Matter* **16**, R755-R769 (2006).

[6] Riggs S. C., Vafek, O., Kemper, J. B., Betts, J. B., Migliori, A., Balakirev, F. F., Hardy, W. N., Liang, R., Bonn, D. A. and Boebinger, G. S. Heat capacity through the magnetic-field-induced resistive transition in an underdoped high-temperature superconductor. *Nature Phys.* **7**, 332-335 (2011).

[7] Doiron-Leyraud, N., Proust, C., LeBoeuf, D., Levallois, J., Bonnemaison, J. -B, Liang, R., Bonn, D. A., Hardy, W. N. & Taillefer, L. Quantum oscillations and the Fermi surface in an underdoped high-$T_c$ superconductor. *Nature* **447** 565-568 (2007).

[8] Sebastian, S. E., Harrison, N., Lonzarich, G. G. Towards resolution of the Fermi surface in underdoped high-$T_c$ superconductors. *Rep. Prog. Phys.* **75**, 102501 (2012).

[9] Harrison, N., Sebastian, S. E. Protected nodal electron pocket from multiple-Q ordering in underdoped high temperature superconductors *Phys. Rev. Lett.* **106**, 226402/1-4 (2011).

[10] LeBoeuf, D., Doiron-Leyraud, N., Daou, R., Bonnemaison, J. B., Levallois, J., Hussey, N. E., Proust, C., Balicas, L., Ramshaw, B., Liang, R., Bonn, D. A., Hardy, W. N., Adachi, S. & Taillefer, L. Electron pockets in the Fermi surface of hole-doped high-$T_c$ superconductors. *Nature* **450**, 533-536 (2007).

[11] Chakravarty, S., Kee, H. -Y. Fermi pockets and quantum oscillations of the Hall coefficient in high-temperature superconductors. *Proc. Nat. Acad. Sci. USA* **105**, 8835-8839 (2008).

[12] Millis A. J., Norman M. R. Antiphase stripe order as the origin of electron pockets observed in 1/8-hole-doped cuprates. *Phys. Rev. B* **76**, 220503(R)/1-4 (2007).

[13] Yao, H., Lee, D. H., Kivelson, S. A. Fermi-surface reconstruction in a smectic phase of a high-temperature superconductor. *Phys. Rev. B* **84**, 012507/1-5 (2011).

[14] Wu, T., Mayaffre, H., Krämer, S., Horvatic, M., Berthier, C., Hardy, W. N., Liang, R., Bonn, D. A. & Julien, M. -H. Magnetic-field-induced charge-stripe order in the high-temperature superconductor YBa$_2$Cu$_3$O$_y$. *Nature* **477**, 191-194 (2011).

[15] Chen, W. -Q., Yang, K. -Y., Rice, T. M., Zhang, F. C. Quantum oscillations in magnetic-field-induced antiferromagnetic phase of underdoped cuprates: Application to ortho-II YBa$_2$Cu$_3$O$_{6.5}$. *Europhys. Lett.* **82**, 17004 (2008).

[16] Sebastian, S. E., Harrison, N., Liang R., Bonn D. A., Hardy W.N., Mielke, C. and Lonzarich, G. G. Quantum oscillations from nodal bilayer magnetic breakdown in the underdoped high temperature superconductor YBa$_2$Cu$_3$O$_{6+x}$ *Phys. Rev. Lett.* **108**, 196403/1-4 (2012).

[17] Sebastian, S. E., Harrison, N., Altarawneh, M. M., Liang, R., Bonn, D. A., Hardy, W. N., and Lonzarich, G. G. Chemical potential oscillations from nodal Fermi surface pocket in the underdoped high-temperature superconductor YBa$_2$Cu$_3$O$_{6+x}$ *Nature Commun.* **2**, 471 (2011).

[18] Hossain, M. A., Mottershead, J. D. F., Fournier, D., Bostwick, A., McChesney, J. L., Rotenberg, E., Liang, R., Hardy, W. N., Sawatzky, G. A., Elfimov, I. S., Bonn, D. A., and Damascelli, A. *In situ* doping control of the surface of high-temperature superconductors. *Nat. Phys.* **4**, 527-531 (2008).

[19] Ghiringhelli G., Le Tacon M., Minola M., Blanco-Canosa S., Mazzoli C., Brookes N. B., De Luca G. M., Frano A., Hawthorn D. G., He F., Loew T., Sala M. M., Peets D. C., Salluzzo M., Schierle E., Sutarto R., Sawatzky G. A., Weschke E., Keimer B., Braicovich L. Long-range incommensurate charge fluctuations in (Y,Nd)Ba$_2$Cu$_3$O$_{6+x}$. *Science* **337**, 821-825 (2012).

[20] Chang, J., Blackburn E., Holmes A. T., Christensen N. B., Larsen J., Mesot J., Liang, R., Bonn D. A., Hardy W. N., Watenphul A., Zimmermann M. v., Forgan E. M. and Hayden S. M. Direct observation of competition between superconductivity and charge density wave order in YBa$_2$Cu$_3$O$_{6.67}$. *Nat. Phys.* **8**, 871-876 (2012).

[21] LeBoeuf, D., Kr$\ddot{a}$mer, S., Hardy, W. N., Liang, R., Bonn, D. A., and Proust, C. Thermodynamic phase diagram of static charge order in underdoped YBa$_2$Cu$_3$O$_y$. *Nature Phys.* **9**, 79-83 (2013).

[22] Hinton, J. P., Koralek, J. D., Lu, Y. M., Vishwanath, A., Orenstein, J., Bonn, D. A., Hardy, W. N., Liang, R. New collective mode in YBa$_2$Cu$_3$O$_{6+x}$ observed by time-domain reflectometry. *Phys. Rev. B* **88**, 060508(R) (2013).

[23] Bergemann, C., Mackenzie, A. P., Julian, S. R., Forsythe, D. and Ohmichi E. Quasi-two-dimensional Fermi liquid properties of the unconventional superconductor Sr$_2$RuO$_4$. *Adv. Phys.* **52**, 639-725 (2003).

[24] Li, J-X., Wu, C-Q., and Lee, D-H. Checkerboard charge density wave and pseudogap of high-$T_\text{c}$ cuprate. *Phys. Rev. B* **74**, 184515 (2006).

[25] Wang, Y., Chubukov, A. V. Charge-density-wave order with momentum $(2Q, 0)$ and $(0, 2Q)$ within the spin-fermion model: Continuous and discrete symmetry breaking, preemptive composite order, and relation to pseudogap in hole-doped cuprates. *Phys. Rev. B* **90**, 035149 (2014).

[26] Castellani, C., Di Castro, C., and Grilli, M. Singular quasiparticle scattering in the proximity of charge instabilities. *Phys. Rev. Lett.* **75**, 4650-4653 (1995).

[27] Hayward, L. E., Hawthorn, D. G., Melko, R. G., Sachdev, S. Angular Fluctuations of a Multicomponent Order Describe the Pseudogap of YBa$_2$Cu$_3$O$_{6+x}$. *Science* **343**, 1336-1339 (2014).

[28] Efetov, K. B., Meier, H., Pépin C. Pseudogap state near a quantum critical point. *Nature Phys.* **9**, 442-446 (2013).

[29] Nie, L., Tarjus, G., Kivelson, S. A., Quenched disorder and vestigial nematicity in the pseudogap regime of the cuprates. *Proc. Nat. Acad. Sci. USA* **111**, 7980-7985 (2014).

[30] Sacuto, A., Gallais, Y., Cazayous, M., M$é$asson, M. -A., Gu, G. D., and Colson, D. New insights into the phase diagram of the copper oxide superconductors from electronic Raman scattering. *Rep. Prog. Phys.* **76**, 022502 (2013).

[31] Allais, A., Chowdhury, D., and Sachdev, S. Connecting high-field quantum oscillations to zero-field electron spectral functions in the underdoped cuprates. *Nature Communications* **5**, 5771 (2014).

# Acknowledgements

S.E.S. acknowledges support from the Royal Society, King's College Cambridge, and the European Research Council under the European Union's Seventh Framework Programme (FP/2007-2013) / ERC Grant Agreement no. 337425-SUPERCONDUCTINGMOTT. N.H. and F.F.B. acknowledge support for high magnetic fields experiments from the the US Department of Energy, Office of Science, BES-MSE 'Science of 100 Tesla' program. G.G.L. acknowledges support from EPSRC grant EP/K012894/1. P.A.G. is supported by the EPSRC and thanks the University of Oxford for the provision of a Visiting Lectureship. R.L., D.A.B., and W.N.H. acknowledge support from the Canadian Institute for Advanced Research, and the Natural Science and Engineering Research Council. A portion of this work was performed at the National High Magnetic

Field Laboratory, which is supported by NSF co-operative agreement no. DMR-0654118, the state of Florida, and the DOE. We acknowledge discussions with many colleagues, including H. Alloul, C. Bergemann, A. Carrington, S. Chakravarty, A. Chubukov, E. M. Forgan, S. R. Julian, B. Keimer, S. A. Kivelson, R. B. Laughlin, M. Le Tacon, L. Taillefer, D. -H. Lee, P. A. Lee, P. B. Littlewood, A. P. Mackenzie, M. R. Norman, C. Pépin, C. Proust, M. Randeria, S. Sachdev, A. Sacuto, T. Senthil, J. P. Sethna, J. Tranquada and C. M. Varma. We are grateful for the experimental support providedby the '100T' team, including J. B. Betts, Y. Coulter, M. Gordon, C. H. Mielke, A. Parish, D. Rickel and D. Roybal.

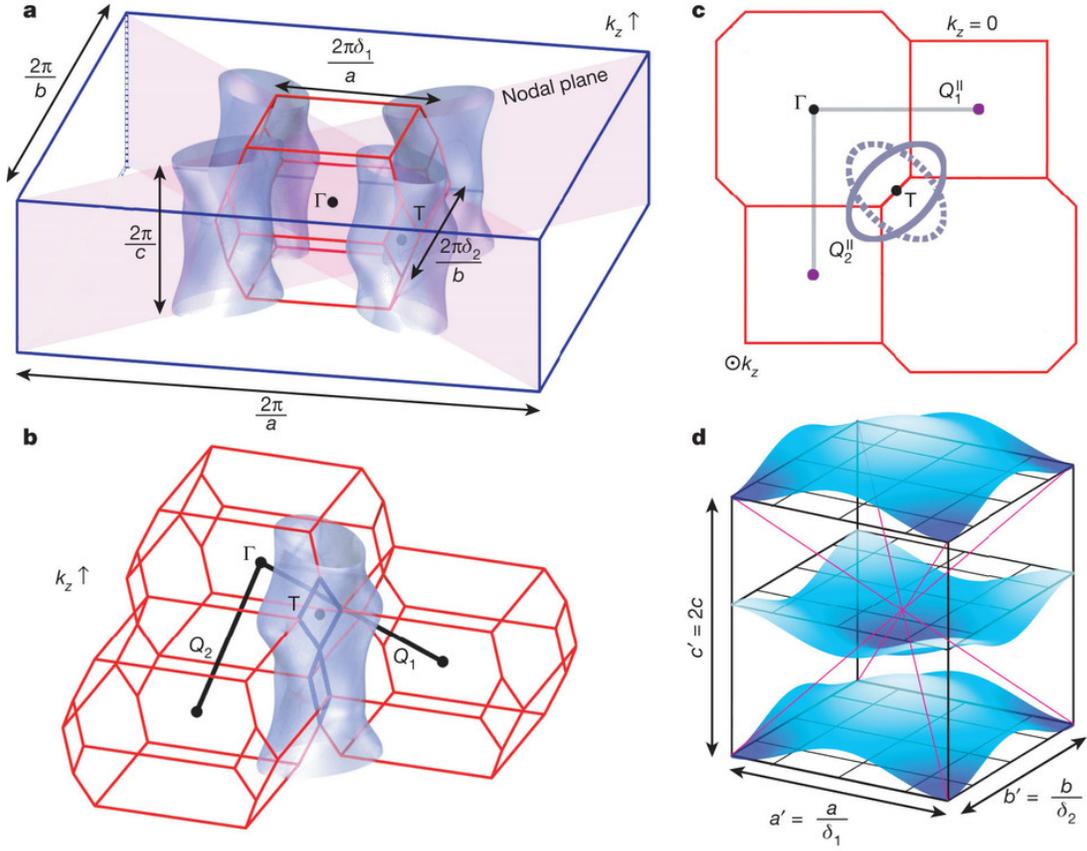

Figure 1: **Fermi surface of $YBa_2Cu_3O_{6.56}$ inferred from quantum oscillation measurements.** **a**, Fermi surface pockets in the nodal planes within the reconstructed body-centred orthorhombic Brillouin zone (in red outlines) and the original primitive orthorhombic Brillouin zone (in blue outlines). **b**, Stacked view of reconstructed Brillouin zones defined by ordering wavevectors $\mathbf{Q}_1 = 2\pi(\pm\frac{\delta_1}{a}, 0, \pm\frac{1}{2c})$ and $\mathbf{Q}_2 = 2\pi(0, \pm\frac{\delta_2}{b}, \pm\frac{1}{2c})$. **c**, Cut of **b** through $k_z = 0$; the dotted line represents the Fermi surface cross-section at a cut through $k_z = \pi/c$. **d**, Blue surfaces represent the inferred sinusoidal superlattice that is staggered between bilayers: here the unprimed and primed lattice parameters relate to the real-space unit cell and the reconstructed superlattice unit cell respectively. $\delta_1$ and $\delta_2$ are given in the text, and the hole doping for $YBa_2Cu_3O_{6.56}$ is approximately 10% holes per Cu atom.

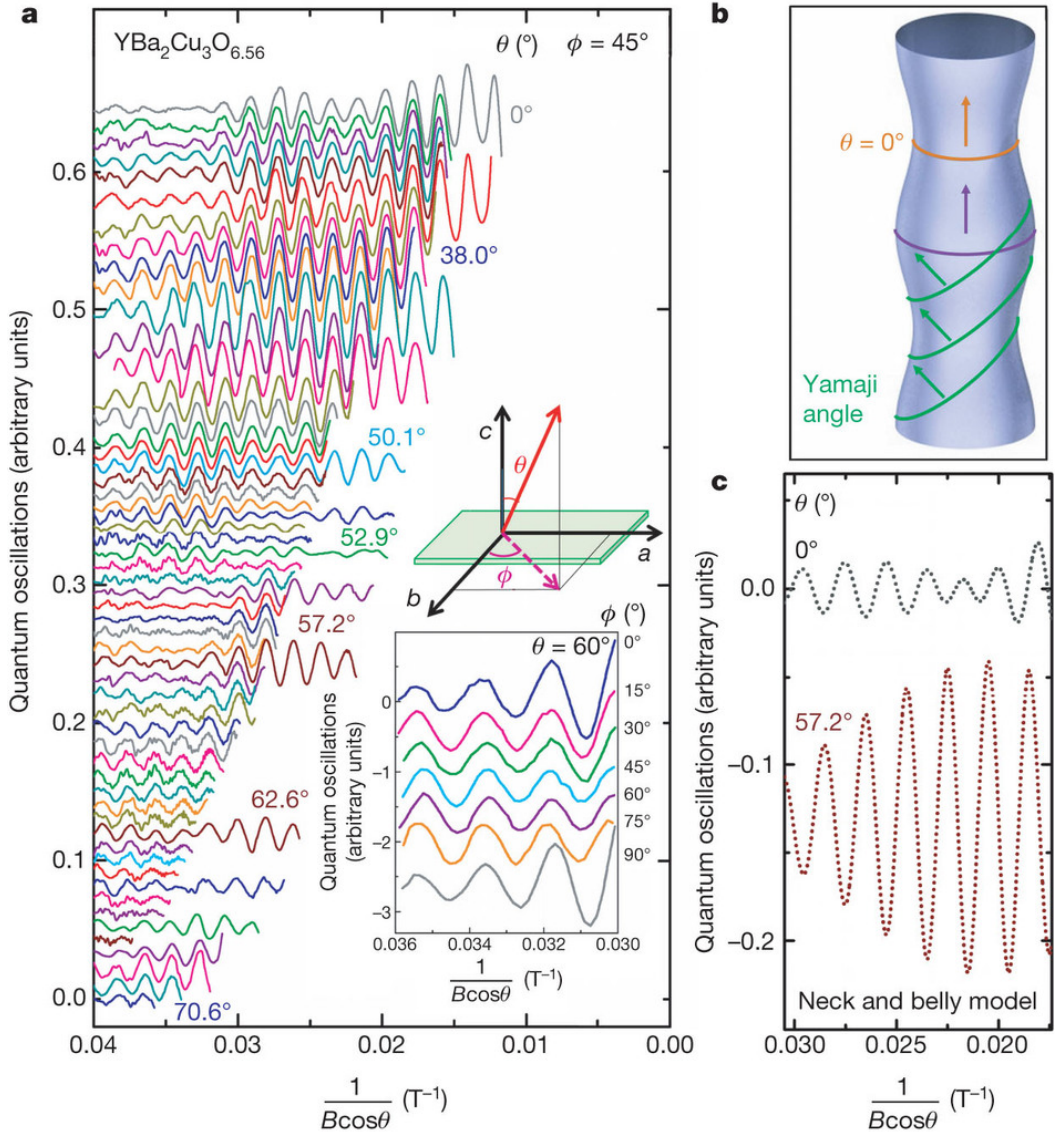

Figure 2: **Quantum oscillations defining the Fermi surface of YBa$_2$Cu$_3$O$_{6.56}$**. **a**, Oscillations of the scaled contactless electrical resistivity as a function of $\frac{1}{B\cos\theta}$ for different magnetic field ($B$) orientations defined by $\phi$ and $\theta$ as indicated. The upper inset defines the angles $\theta$ and $\phi$ relative to the crystallographic axes. **b**, Schematic showing the degeneracy in the cyclotron orbit cross-sectional area that yields an amplitude enhancement at the Yamaji angle (green) for a neck and belly Fermi surface geometry (corrugation is accentuated throughout for clarity). **c**, Expected Yamaji resonance in the vicinity of $\approx 60°$ were YBa$_2$Cu$_3$O$_{6.56}$ to be described by a neck and belly Fermi surface geometry (see Methods, Extended Data Fig. 2**b**,**c**). Data in **a** and simulations in **c** have been scaled by $\exp\frac{100\ \mathrm{T}}{B\cos\theta}$ for visual clarity.

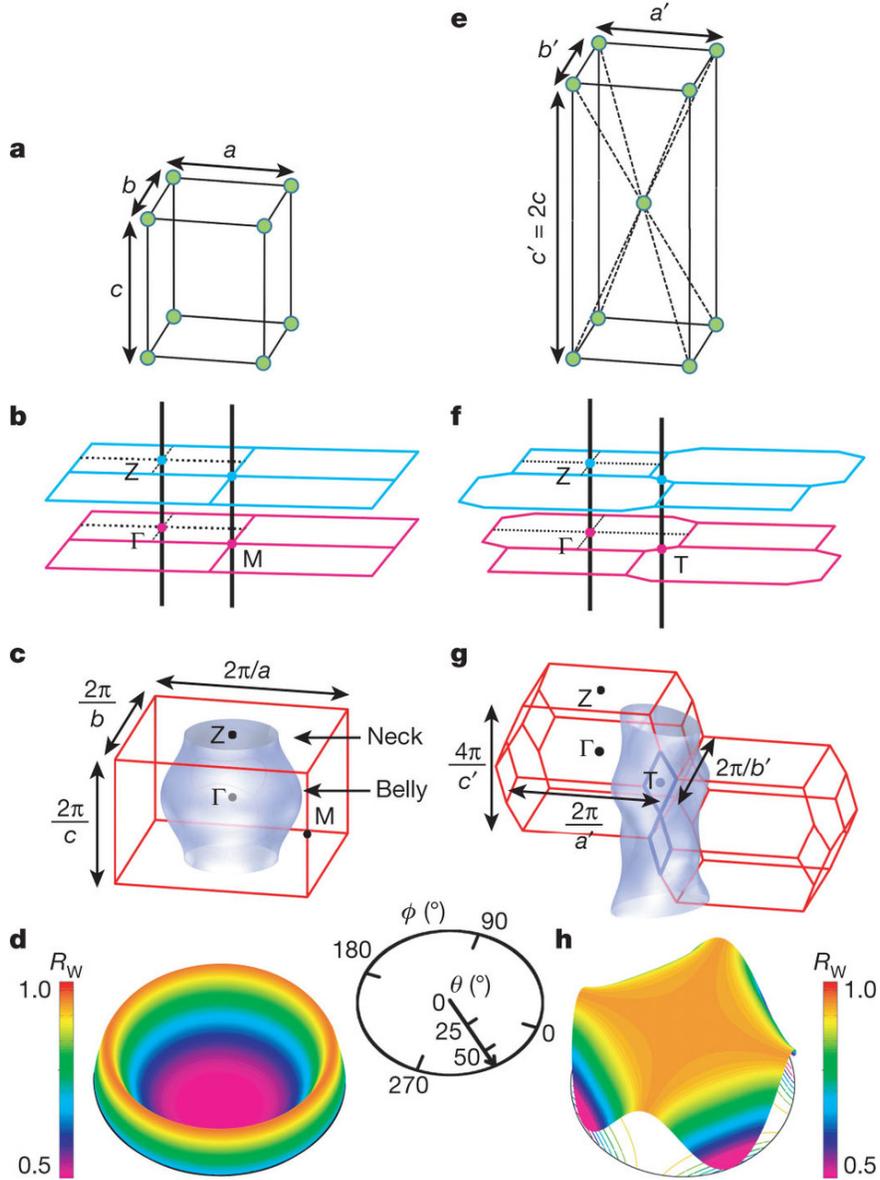

Figure 3: **Fermi surface and geometry-dependent quantum oscillation damping for different crystal structures**. **a-d**, Primitive orthorhombic structure. **e-h**, Body-centered orthorhombic structure. Shown are the real space unit cells (**a**, **e**), cuts through the Brillouin zone showing local symmetries about the vertical lines (**b**, **f**), location of the warped Fermi surface within the Brillouin zone in which symmetry points $\Gamma$, $Z$, $M$, and $T$ are indicated (**c**, **g**), and the angular dependences of the corresponding quantum oscillation amplitude damping factor $R_\text{w}$ (**d**, **h**). The amplitude for the fundamental warping (**d**) exhibits a maximum versus $\theta$ − known as the Yamaji resonance − in contrast to an anti-resonance for the staggered twofold warping (**h**). $R_\text{w}$ has been simulated using parameters in Extended Data Table 1 and a representative value of $B = 62$ T. For the definition of $R_\text{w}$ and further analyses including the spin factor $R_\text{s}$ see the Methods.

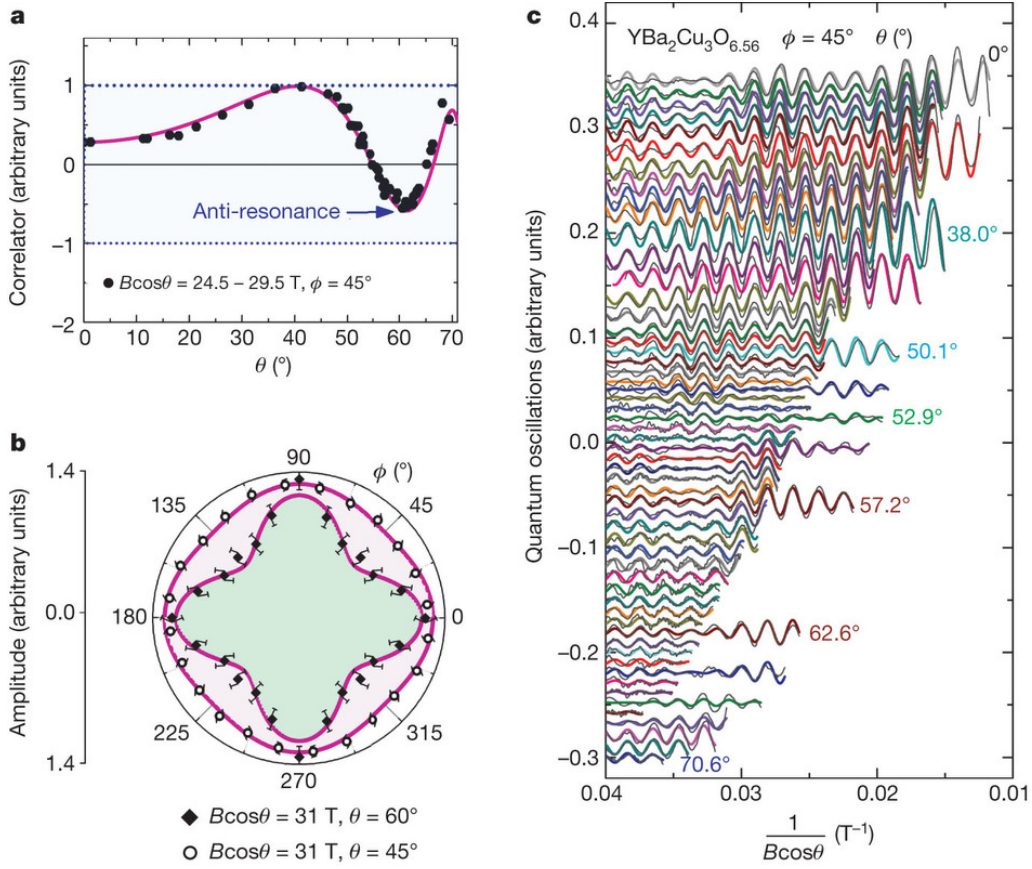

Figure 4: **Quantum oscillation data compared with a staggered twofold Fermi surface model. a**, Real component of the cross-correlation between the quantum oscillation data over a fixed range of $B\cos\theta$ for a range of measured $\theta$ angles with a simple sinusoid $\cos(2\pi F/B\cos\theta + \phi)$. $F$ and $\phi$ are matched to the periodicity and phase of the oscillations at $\theta = 38°$, where a single frequency dominates the measured quantum oscillations. The dotted lines show the expected constant magnitude of the maximum and minimum amplitude for an ideal two-dimensional Fermi surface. A suppression in amplitude (anti-resonance) instead of a Yamaji resonance is observed as a function of $\theta$ (see Methods, Extended Data Fig. 2). **b**, Fourfold anisotropy in the $\phi$-dependent amplitude of quantum oscillations for $\theta = 45°$ and $60°$. **c**, Measured quantum oscillations shown in Fig. 2 (black lines). The magenta lines in **a** and **b** show a simulation of $R_\mathrm{w}^\mathrm{twofold} R_\mathrm{s}$ and the coloured lines in **c** show a simulation of $\Psi_\mathrm{twofold}$ respectively for a staggered twofold Fermi surface model (for a value of $\Delta F_\mathrm{twofold} = 15$ T). For definitions of the staggered twofold damping factor $R_\mathrm{w}^\mathrm{twofold}$, the spin damping factor $R_\mathrm{s}$, the quantum oscillation function $\Psi_\mathrm{twofold}$, the modulation amplitude of staggered twofold geometry $\Delta F_\mathrm{twofold}$, and parameter values used for the simulation, see the Methods.

# Methods

## Experimental Details.

High quality detwinned single crystals of YBa$_2$Cu$_3$O$_{6.56}$ of dimensions 0.5 x 0.8 x 0.1 mm$^3$ used in this study were grown and prepared at the University of British Columbia.[31] Quantum oscillations were measured at the National High Magnetic Field Laboratory (Los Alamos and Tallahassee) using the contactless resistivity technique (described elsewhere[32,33]). The sample and the proximity detector oscillator coil to which it is coupled were rotated *in situ* through different polar ($\theta$) and azimuthal ($\phi$) angles before each magnetic field pulse extending to 65 or 85 T. Quantum oscillation measurements for $\theta = 0$ were extended to 100 T, in the regime where long range superconductivity is destroyed at low temperatures.[2,3,34,35] Additional azimuthal angle dependences were measured using a two-axis goniometer in the 45 T DC hybrid magnet. In all cases the sample temperature was maintained close to $T \approx 1.5$ K throughout the experiment by direct immersion in superfluid $^4$He. A worm drive rotator powered by a stepper motor at room temperatures was used for sample rotation, with a secondary angular calibration provided by a pancake projection coil wound in the plane of the sample. For the pulsed-field data, the measured component of the applied magnetic induction $B \cos \theta$ (where $B \approx \mu_0 H$) projected along the crystalline $c$-axis shown in Extended Data Fig. 1 yields an uncertainty of approximately $0.2°$ or less in the sample orientation.

The polar angles accessed in the rotation study shown in Figs. 2**c** and 4**a** and Extended Data Figs. 1 and 2 are $\theta = 0°$, 1.3°, 11.3°, 12°, 16.3°, 18°, 21.3°, 26.3°, 31.3°, 36.3°, 38°, 41.3°, 45.2°, 46.3°, 48°, 49°, 49.4°, 50.1°, 50.6°, 51.4°, 51.5°, 52°, 52.3°, 52.5°, 52.9°, 53.1°, 54.4°, 54.9°, 55.5°, 56°, 56.2°, -56.95°, 57.2°, -57.4°, -58.15°, 58.2°, -59.4°, 59.6°, 60.6°, 61.2°, -61.4°, 61.7°, 62.5°, 62.6°, -62.7°, -63.2°, 63.4°, 63.7°, -64.1°, 64.5°, 65.5°, 66°, 66.3°, 68.1°, 69.4° and 70.6°. Negative $\theta$ angles refer to a measured equivalent (180-$|\theta|$) angle, as shown in

Extended Data Fig. 1.

## Absence of experimental Yamaji effect in underdoped YBa$_2$Cu$_3$O$_{6+x}$.

Extended Data Fig. 2**a** shows a compilation of quantum oscillations measured for YBa$_2$Cu$_3$O$_{6.56}$ (also shown in Figs. 2**a** and 4**c**). In previous experiments over a limited range of magnetic field and orientation (indicated by thick grey dashed line in Extended Data Fig. 2**b**),[36] experimental data was compared with a simulated quantum oscillation waveform from two cylinders with fundamental neck and belly geometry:

$$\Psi_{\text{neck-belly}} = \sum_j a_j [R_s R_D R_T R_w]_j \sin\left(\frac{2\pi F_j}{B\cos\theta} - 2\pi\gamma_j\right), \quad (1)$$

where $j = 1, 2$ for two cylinders, and the relative amplitudes $a_j$, phases $\gamma_j$, quantum oscillation frequencies $F_j$, and fundamental neck and belly warpings $\Delta F_{\text{neck-belly},j}$ were considered as independent parameters.[36] The spin damping factor $R_s$, Dingle damping factor $R_D$, thermal damping factor $R_T$, and the geometrical damping, or warping factor $R_w$ are defined in the following sections. In this case, the quantum oscillatory frequency spread is modelled by neck and belly warping, which predicts a dramatic enhancement in quantum oscillation amplitude at a special Yamaji polar angle of approximately 60 ◦. This Yamaji resonance is a consequence of all the orbits becoming degenerate at this special angle.[23,37] Simulations made using parameters from ref.[36] (shown in Extended Data Table 3) are shown in Fig. 2**b** and **c** and Extended Data Fig. 2**b** and **c**. In the present experiment in underdoped YBa$_2$Cu$_3$O$_{6+x}$, where an extended magnetic field and polar and azimuthal angular range are accessed, despite the sizeable frequency spread, we unexpectedly find the absence of any quantum oscillation amplitude enhancement in the vicinity of a polar Yamaji angle of approximately 60°, with instead a suppression of the quantum oscillation amplitude at a special polar angle in the vicinity of ≈ 60°.

## Simulations using staggered twofold Fermi surface geometry.

Fig. 4 shows a simulation of a Fermi surface with staggered twofold geometry that agrees with the experimental quantum oscillation waveform and amplitude remarkably well over the entire experimental range in $B$, $\theta$ and $\phi$. The simulation uses a quantum oscillation function of the form expected for a bilayer-split nodal Fermi surface with staggered twofold warping (see Fig. 1**a**-**c**)

$$\Psi_{\text{twofold}} = a_0 \sum_{j=1}^{6} N_j [R_{\text{w}} R_{\text{MB}} R_{\text{s}} R_{\text{D}} R_T]_j \cos\left(\frac{2\pi F_j}{B\cos\theta} - \pi\right). \tag{2}$$

Here, $a_0$ is the amplitude prefactor (which is taken to be the same for all orbits), $R_{\text{MB}}$ is the magnetic breakdown amplitude reduction factor to be defined in the following sections, and $N_j$ counts the number of instances the same orbit is repeated within the magnetic breakdown network. In our model, $R_{\text{w}}$, $R_{\text{D}}$, and $R_T$ are taken as the same for all orbits.

## Conventional damping parameters included in simulations.

Quantum oscillation simulations include conventional thermal, Dingle, and spin damping factors of the same form used for previous comparisons with quantum oscillations measured in the underdoped copper oxides and other layered families of materials.[7,8,10,16,17,32,36,38–53] The thermal damping factor is given by

$$R_T = \frac{X_j}{\sinh X_j}$$

(where $X_j = 2\pi^2 k_{\text{B}} m^*_{\theta j} T/\hbar e B$), and the Dingle damping factor is given by

$$R_{\text{D}} = \exp\left(-\frac{\Lambda_j}{B\cos\theta}\right),$$

(where $\Lambda_j$ is a damping factor).[38] The spin damping factor is given by

$$R_{\text{s}} = \cos\left[\frac{\pi}{2}\left(\frac{m^*_{\theta j}}{m_{\text{e}}}\right) g^*_{\theta j}\right],$$

where $m^*_{\theta j} = m^*_{\|j}/\cos\theta$ for a given orbit '$j$' is determined by the projection $B\cos\theta$ of $B$ perpendicular to the planes (that is, the projection parallel to the $\hat{c}$-axis in YBa$_2$Cu$_3$O$_{6+x}$). The anisotropic effective $g$-factor has the form $g^*_{\theta j} = g^*_{\|j}\sqrt{\cos^2\theta + \frac{1}{\xi_j}\sin^2\theta}$, which can be renormalised with respect to the electron $g$-factor by spin orbit coupling,[54] many body effects[55] and the presence of small band gaps.[56,57] Here, $m^*_{\|j}$ and $g^*_{\|j}$ refer to the respective values of $m^*_{\theta j}$ and $g^*_{\theta_j}$ when $B$ is parallel to the crystalline $\hat{c}$-axis, while $\xi_j = \left(\frac{g^*_\|}{g^*_\perp}\right)^2$ is the anisotropy in the spin susceptibility. Because of the multiple frequencies in the model, it is not possible to identify the $g$-factors uniquely. The values of $g^*_j$ and $\xi^*_j$ here represent parameters used for the simulation.

The quantum oscillatory frequency spread is modelled by a splitting of the Fermi surface, which can arise not only by tunnelling between bilayers,[16,58] but also, for example, via spin-orbit effects under certain conditions.[59] This leads to two starting frequencies that are denoted for reasons that will become clear below, as $F_0 - 2F_{\text{split}}$ and $F_0 + 2F_{\text{split}}$.[16] Magnetic breakdown tunneling (in the nodal region where the splitting is smallest) can then give rise to a series of combination frequencies. Here we consider cyclotron orbits corresponding to a nodal bilayer-split Fermi surface from charge ordering shown in Extended Data Fig. 6 and 7 (modelled in refs.,[8,9,16,60,61]). The combination frequencies from the Fermi surface orbits depicted in Extended Data Fig. 7 are listed in Extended Data Table 2. For a small magnetic breakdown gap, a high tunnelling probability $P$ causes the orbit of frequency $F_0$ to dominate the quantum oscillation amplitude (as seen in Extended Data Fig. 3a). The beat structure in the quantum oscillation simulation is caused primarily by the superposition of the dominant $F_0$ oscillations and weaker amplitude oscillations of frequency $F_0 - \Delta F_{\text{split}}$ and $F_0 + \Delta F_{\text{split}}$. Oscillations of frequency $F_0 - 2\Delta F_{\text{split}}$ and $F_0 + 2\Delta F_{\text{split}}$ are expected to be even more strongly attenuated. The magnetic breakdown amplitude reduction factor has the form

$$R_{\text{MB}} = (i\sqrt{P})^{l_\nu}(\sqrt{1-P})^{l_\eta},$$

in which $l_\nu$ and $l_\eta$ count the number of magnetic breakdown tunnelling and Bragg reflection events *en route* around the orbit, having transmitted amplitudes $i\sqrt{P}$ and $\sqrt{1-P}$ respectively. The magnetic breakdown probability is given by $P = \exp(-B_0/B\cos\theta)$, where $B_0$ is the characteristic magnetic breakdown field.[38]

**Angle-dependent damping term from Fermi surface warping geometry.**

The Fermi surface warping geometry in layered materials leads to a quantum oscillation damping factor with an azimuthal ($\phi$) and polar ($\theta$) angular variation that is very sensitive to the energy band dispersion perpendicular to the planes (known as the 'warping' of the Fermi surface). The amplitude damping factor ($R_\mathrm{w}$) which arises as a consequence of weak Fermi surface warping is obtained by (1) an expansion of the Fermi surface wavevectors in terms of cylindrical harmonics, and retaining only the lowest order harmonic in $k_z$, (2) an evaluation of the Fermi surface cross-sectional area $A(k_z, \theta, \phi)$ in a plane normal to the field orientation defined by $(\theta, \phi)$ and crossing $k_z$, and (3) via a $k_z$ integration of $\exp(i\hbar A(k_z, \theta, \phi)/(eB))$.

For the case of a primitive weakly-orthorhombic unit cell[62] (with lattice dimension $c$, see Fig. 3**a**) in the limit of weak warping, we thus arrive at the quantum oscillation damping factor for leading order neck and belly geometry, in the simplified form

$$R_\mathrm{w}^{\mathrm{neck-belly}} = J_0\left[\frac{2\pi\Delta F_0}{B\cos\theta}J_0(k_F c\tan\theta)\right] \qquad (3)$$

In contrast, for the case of a body-centred weakly-orthorhombic unit cell (with lattice dimension $c' = 2c$, see Fig. 1**d** and 3**e**) in the limit of weak warping, the quantum oscillation damping factor for leading order staggered twofold geometry has the simplified form

$$R_\mathrm{w}^{\mathrm{twofold}} = J_0\left[\frac{2\pi\Delta F_1}{B\cos\theta}\sin 2\phi J_2(k_F(c'/2)\tan\theta)\right] \qquad (4)$$

The former has well known examples occurring in materials families including organic conductors[39–41] while the latter occurs at the Brillouin zone corner in materials families including body-centered tetragonal pnictides,[51] ruthenates[23] and Tl-based (overdoped) high $T_\mathrm{c}$ superconductors.[49,52]

Here the effective Fermi wavevector $k_\mathrm{F}$ (the average radius of the Fermi pocket in the basal plane) and warping parameters $\Delta F_n$ are related to the cylindrical harmonic expansion parameter $k_{\mu\nu}$ defined in ref.[23] by $k_\mathrm{F} = k_{00}$ and $\Delta F_n = \hbar k_{00} k_{2n,1}/e$, as is appropriate to the leading order, $\nu = 1$, cylindrical harmonic in $k_z$. In this work, we refer to the amplitude of fundamental neck and belly geometry $\Delta F_0$ as $\Delta F_\mathrm{neck-belly}$, and the amplitude of staggered twofold geometry $\Delta F_1$ as $\Delta F_\mathrm{twofold}$. The relevant Fermi surface geometry depends on the local symmetry in the Brillouin zone. Examples of an isotropic azimuthal anisotropy for a fundamental neck and belly Fermi surface geometry and a fourfold azimuthal anisotropy for a staggered twofold Fermi surface geometry are shown in Fig. 3.

While a fundamental neck and belly geometry would be expected to dominate at all locations in a primitive orthorhombic Brillouin zone, a staggered twofold Fermi surface geometry would arise from the unique symmetry of the special corner T point in the Brillouin zone of a **body-centred** orthorhombic unit cell (shown in Figs. 3**f,g**, and 1**a-c**). At this corner T point, the twofold in-plane rotational symmetry is seen to be rotated by 90° between adjacent symmetry planes. The special corner point is characterised by a leading order expansion of the Fermi wavenumber

$$k_\mathrm{F}(k_z, \phi) = k_\mathrm{F}(\phi) + \Delta k_\mathrm{twofold} \cos((c'/2)k_z) \sin(2\phi), \qquad (5)$$

in cylindrical harmonics, where $k_z$ is the wavenumber along the $\hat{c}$-axis, $c' = 2c$ (Fig. 1**d** and 3**e**), and the parameter $\Delta k_\mathrm{twofold}$ is the amplitude of modulation along the $\hat{c}$-axis, with the following symmetries: (1) invariance under the joint transformation $k_z \to k_z + 2\pi/c'$ and $\phi \to \phi + 90°$, (2) mirror symmetry about $k_z = 0$ and (3) mirror symmetry about $\phi = 45°$ (that is, the nodal planes in $YBa_2Cu_3O_{6+x}$ shown in Fig. 1**a**). While this special symmetry of the corner point does not support a neck and belly geometry, it instead supports a Fermi surface with a stag-gered twofold geometry (Figs. 1**a** to **c** and 3**g**, and examples from other layered unconventional superconductors[23,51,52]).

We see from the simulated geometrical quantum oscillation amplitude damping factor as a function of polar and azimuthal angle that a fundamental neck and belly geometry and a staggered twofold Fermi surface geometry would exhibit strikingly different angular dependences (shown in Figs. 3**a**-**d** and 3**e**-**h** respectively, simulated using equations 3 and 4). A staggered twofold Fermi surface geometry would yield an amplitude suppression (anti-resonance) in quantum oscillation amplitude at a special polar angle, in contrast to the enhancement in quantum oscillation amplitude expected at a special Yamaji polar angle[23,37] for a fundamental neck and belly Fermi surface geometry. Furthermore, while a fundamental neck and belly Fermi surface would yield an isotropic azimuthal dependence of quantum oscillation amplitude to leading order, a staggered twofold Fermi surface geometry would yield a fourfold azimuthal dependence of quantum oscillation amplitude to leading order.

## Staggered twofold Fermi surface geometry simulation.

Extended Data Table 1 shows the value of parameters used for the quantum oscillation waveform simulated for a staggered twofold Fermi surface using Eqns. 2 and 4, which is compared with the experimental data in Fig. 4 and Extended Data Figs. 2**c** and 4. For simplicity, the value of magnetic breakdown field $B_0$, warping $\Delta F_{\text{twofold}}$, damping $\Lambda$ and effective mass $m^*_\parallel$ is taken to be the same for all orbits (enabling us to drop the subscript $j$). Furthermore, only two sets of anisotropic $g$-factors are considered: orbits $F_1$, $F_2$, $F_4$, $F_5$, and $F_6$, which undergo both magnetic breakdown tunnelling and finite Bragg reflection (that is, $1 - P$) are approximated to have the same $g$-factor $g^*_{\parallel\diamond}$ with anisotropy $\xi_\diamond$. This anisotropy is smaller than for orbits $F_3$, which show only magnetic breakdown tunnelling without finite Bragg reflection, and which are approximated to have a common $g$-factor $g^*_{\parallel\square}$ with anisotropy $\xi_\square$. Since it is not possible to identify unique $g$-factor values given the multiplicity of frequencies, the values of $g_j^*$ and $\xi_j^*$ here represent parameters used for the simulation.

The amplitudes of the $F_2$ and $F_5$ orbits have approximated to be equal, while experimentally, the amplitude of the $F_2$ orbit is slightly larger in contactless resistivity[16] and magnetic torque[44] measurements, probably due to additional effects such as small differences in the scattering time or effective mass. We note that the amplitude of the $F_2$ orbit is significantly larger in $c$-axis transport experiments.[36]

We note that in Eqns. 2, 4, 5 the most general circular Fermi surface cross-section has been assumed, which is not specific to any Fermi surface reconstruction model. In order to extract the in-plane Fermi surface topology from the experimental data, a complete determination of the $k_{\mu,\nu}$ Fermi wavevectors would be required, for which future complementary experiments to the present quantum oscillation experiments are indicated.

## Additional angle-dependent damping factors considered.

The inclusion of damping of the quantum oscillation amplitude from a finite lifetime of the quasiparticles would further not be expected to alter the conclusion of a staggered twofold Fermi surface geometry. This damping can be described in terms of a complex dispersion relation or effectively in terms of complex Fermi wavevectors, which are consistent with the symmetry of the full lattice potential both in their real and imaginary parts. Here the effect of quasiparticle scattering lifetime over the Fermi surface is included in terms of an imaginary component of the Fermi wavevector. The anisotropy of the amplitude arising from a complex Fermi surface (defined by Fermi wavevectors with real and imaginary parts) is expected to be the same as the original real Fermi surface. If the imaginary part of the warping were to dominate the real part then the amplitude damping factors would be given by expressions similar to Eqns. 3 and 4 to leading order in warping, but with the outer Bessel function $J_0$ replaced by the modified

Bessel function $I_0$, and $\Delta F_0$ and $\Delta F_1$ replaced by analogous parameters that measure the degree of damping. Our conclusion wherein the quantum oscillatory frequency spread is inconsistent with a fundamental neck and belly warping remains unchanged for the reasons that (1) for a real part of the warping sizeable enough to capture the frequency spread, a Yamaji resonance would occur, contrary to observation, whereas (2) if the real part were negligible then the amplitude would be governed by the very different function $I_0$, which is also contrary to experimental observation.

Other amplitude damping factors that arise from random quenched sample inhomogeneities or magnetic field inhomogeneities would not be expected to show the symmetries of the underlying lattice. Further damping factors such as additional damping within the vortex regime, or a damping factor of the form $\exp(-1/(\cos\theta)^{\alpha-1})$ where $\alpha$ is a variable parameter (after ref.[49]) have been considered. However, such damping factors would yield an increasing amplitude suppression with increase in polar angle, and hence do not provide an explanation for the observed anti-resonance in quantum oscillation amplitude in the vicinity of a special polar angle $\approx 60°$ in underdoped $YBa_2Cu_3O_{6.56}$.

## Negative Hall effect from nodal Fermi surface pocket from charge order.

Of the various models proposed for the normal ground state of the underdoped copper oxides,[1,4,5,7–15,24–29,48,60,61,63–92] examples of Fermi surface models that have been proposed include refs.[6–13,15–17,36,48,53,60,61,63,93–100] An accurate Fermi surface model would need to explain the observation of quantum oscillations in underdoped $YBa_2Cu_3O_{6+x}$ on a background of a negative Hall coefficient $R_H$ (ref.[10]), given that a positive Hall coefficient would be expected for the hole carrier doping in underdoped $YBa_2Cu_3O_{6+x}$. In various previously proposed interpretations, the negative Hall effect is yielded by electron pockets at the antinodal locations of the Brillouin zone.[8,11–13,53] Our present work that locates the small Fermi pockets in the vicinity of the nodal regions of the Brillouin zone, however, renders these interpretations unlikely.

The nodal staggered twofold Fermi pocket in the new reconstructed body-centred orthorhombic Brillouin zone that we propose to arise from staggered charge order (schematic in Extended Data Figs. 6 and 7; refs.[8,9,16,60,61]), has opposite polarity to the original large Fermi surface in the starting primitive orthorhombic Brillouin zone in underdoped $YBa_2Cu_3O_{6+x}$. In the high magnetic field limit, the Hall conductivity from a Fermi surface pocket − rather than being determined by the Fermi surface curvature, as would be the case in low magnetic fields[101] − would instead be expected to be proportional to the number of states contained within that pocket, with a sign that is negative for filled states and positive for empty states.[102,103] A negative Hall coefficient would therefore arise from the nodal staggered twofold Fermi surface from charge order,[8,9,16,60,61] providing an explanation for the previous puzzling observation of quantum oscillations in the negative Hall coefficient.[10]

## Low quantum oscillation frequency.

Very slow quantum oscillations are further seen over a large magnetic field range in contactless electrical resistivity measurements (Extended Data Fig. 3b). The separation between observed oscillation maxima (lower inset to Extended Data Fig. 3b) corresponds to a frequency of $90 \pm 10$ T.

A property of systems exhibiting magnetic breakdown is the occurrence of Stark quantum interference effects in electrical transport.[104,105] Quasiparticles that encounter the magnetic breakdown junction (indicated in magenta in the upper inset of Extended Data Fig. 3) can take two possible paths (i.e. the inner orbit or the outer orbit), depending on whether or not magnetic breakdown tunnelling occurs. When the quasiparticles recombine at a second magnetic breakdown junction they acquire a relative difference in Onsager phase ($\Delta\phi = 2\pi\Delta F/B$), which is proportional to the difference in $k$-space area between the two paths. Because this area is not a closed orbit, it does not yield quantum oscillations in thermodynamic quantities such as the magnetisation or heat capacity − however, interference between the two paths can give rise to oscillations in electronic transport properties such as the electrical resistivity, or thermal conductivity. The simulation of the charge order model made using parameters in External Data Tables 1 and 2 yields $\Delta F \approx 90$ T for the frequency corresponding to the small difference in area between the two magnetic breakdown junctions, suggesting the association of the observed low quantum oscillation frequency with the Stark quantum interference effect.[104] Alternatively, this low frequency could correspond to the small hole pocket predicted in ref.[31] The existence of a small hole pocket in addition to the nodal electron pocket discussed in the main text could provide a possible explanation for the observed oscillatory component of the Hall resistivity.[?,7]

**Anisotropy between in-plane and inter-plane effective hopping.**

The velocity anisotropy corresponding to a staggered twofold Fermi surface geometry may be estimated by means of a quasi-two dimensional energy band dispersion of a form consistent with the order of expansion in cylindrical harmonics in equation 4. In cylindrical coordinates $(k, \phi, k_z)$ and to an additive constant, this dispersion relation near the Fermi surface can be expressed in the form

$$\varepsilon = \frac{\hbar^2 k^2}{2m^*} + 2t^* \cos k_z(c'/2) \sin 2\phi,$$

where the magnitude of the interlayer dispersion varies around the circumference. Here $m^*$ is an in-plane effective mass, $k_\mathrm{F}$ is the effective Fermi wavevector, and $t^*$ is a $c$-axis effective hopping parameter.

The ratio of the in-phase Fermi velocity $v_\parallel$ to $c$-axis root mean square velocity $v_\perp$ is then

$$\frac{v_\parallel}{v_\perp} = \Big(\frac{F_0}{\Delta F_\mathrm{twofold}}\Big)\Big(\frac{4}{k_\mathrm{F}c}\Big)$$

where $\Delta F_\mathrm{twofold}/F_0 = 2t^*/\big(\hbar^2 k_\mathrm{F}^2/2m^*\big)$. Taking $F_0 \approx 530$ T, $\Delta F_\mathrm{twofold} \approx 15$ T, $k_\mathrm{F} = \sqrt{2eF_0/\hbar} \approx 0.13$ Å$^{-1}$ and $c \approx 11.7$Å, the velocity anisotropy ratio $v_\parallel/v_\perp$ is of the order of 100, a large value consistent with a nodal location of the Fermi surface pocket, given that the interlayer hopping is expected to be weakest in this region of the Brillouin zone.[62] Interestingly, the velocity anisotropy ratio is of a similar order of magnitude as the conductivity anisotropy ratio inferred for optical conductivity measurements in the pseudogap regime.[106]

The correlation length of the associated superlattice structure[19,20] is required to be similar to or greater than the extent of motion of the cyclotron orbit in the interlayer direction, for our analysis arriving at the Fermi surface geometry in this work. The large anisotropy in velocity between in-plane and interlayer directions means that the cyclotron orbits in real space traverse an interlayer distance of only about 1 % of the in-plane cyclotron orbit radius, placing us in the required limit.

# References and Notes

[31] Liang R., Bonn D. A., Hardy, W. N., Evaluation of CuO$_2$ plane hole doping in YBa$_2$Cu$_3$O$_{6+x}$ single crystals *Phys. Rev. B* **73**, 180505 (2006).

[32] Sebastian, S. E. *et al.*, Metal-insulator quantum critical point beneath the high Tc superconducting dome *Proc. Nat. Acad. Sci. USA* **107**, 6175-6179 (2010).

[33] Altarawneh, M. M., Mielke, C. H., Brooks, J. S. Proximity detector circuits: an alternative to tunnel diode oscillators for contactless measurements in pulsed magnetic field environments. *Rev. Sci. Instr.* **80**, 066104 (2009).

[34] Ramshaw, B. J., Day, J., Vignolle, B., LeBoeuf, D., Dosanjh, P., Proust, C., Taillefer, L., Liang, R., Hardy, W. N., Bonn, D. A. Vortex Lattice Melting and $H_{c2}$ in underdoped YBa$_2$Cu$_3$O$_y$ *Phys. Rev. B* **86**, 174501/1-7 (2012).

[35] Ruiller-Albenque, F., Alloul, H., Colson, D., Forget, A. Determination of superconducting fluctuations in high-$T_\mathrm{c}$ cuprates *J. Phys.: Conf. Series* **449** 012010/1-10 (2013).

[36] Ramshaw, B. J., Vignolle, B., Day, J., Liang, R., Hardy, W. N., Proust, C., and Bonn, D. A. Angle dependence of quantum oscillations in YBa$_2$Cu$_3$O$_{6.59}$ shows free-spin behaviour of quasiparticles *Nature Phys.* **7**, 234-238 (2011).

[37] Yamaji, K. On the angle dependence of the magnetoresistance in quasi-two-dimensional organic superconductors *J. Phys. Soc. Jpn* **58**, 1520-1523 (1989).

[38] Shoenberg D., *Magnetic oscillations in metals*, (Cambridge University Press, Cambridge 1984).

[39] Singleton, J., Shoenberg D., *Band Theory and Electronic Properties of Solids*, (Oxford University Press, Oxford 2001).

[40] Wosnitza, J. Fermi surfaces of organic superconductors. Int. J. Mod. Phys. B **7**, 2707-2741 (1993).

[41] Kartsovnik, M. V. High magnetic fields: A tool for studying electronic properties of layered organic metals *Chem. Rev.* **104**, 5737-5781 (2004).

[42] Yelland, E. A., Singleton, J., Mielke, C. H., Harrison, N., Balakirev, F. F., Dabrowski, B., Cooper, J. R. Quantum Oscillations in the Underdoped Cuprate YBa$_2$Cu$_4$O$_8$. *Phys. Rev. Lett.* **100**, 047003/1-4 (2008).

[43] Bangura A. F., Fletcher, J. D., Carrington, A., Levallois, J., Nardone, M., Vignolle, B., Heard, P. J., Doiron-Leyraud, N., LeBoeuf, D., Taillefer, L., Adachi, S., Proust, C., Hussey, N. E. Quantum Oscillations in the Underdoped Cuprate YBa$_2$Cu$_4$O$_8$ *Phys. Rev. Lett.* **100**, 046004/1-4 (2008).

[44] Audouard, A. *et al.* Multiple Quantum Oscillations in the de Haas−van Alphen Spectra of the Underdoped High-Temperature Superconductor $YBa_2Cu_3O_{6.5}$ *Phys. Rev. Lett.* **103**, 157003/1-4 (2009).

[45] Sebastian, S. E., Harrison, N., Palm, E., Murphy, T. P., Mielke, C. H., Liang, R., Bonn D. A., Hardy, W. N., and Lonzarich, G. G. A multi-component Fermi surface in the vortex state of an underdoped high-$T_c$ superconductor. *Nature* **454**, 200-203 (2008).

[46] Singleton, J., de la Cruz, C., McDonald, R. D., Li, S., Altarawneh, M., Goddard, P. A., Franke, I., Rickel, D., Mielke, C. H., Yao, X., Dai, P. Magnetic quantum oscillations in $YBa_2Cu_3O_{6.61}$ and $YBa_2Cu_3O_{6.69}$ in fields of up to 85 T: patching the hole in the roof of the superconducting dome. *Phys Rev Lett.* **104**, 086403/1-4 (2010).

[47] Sebastian, S. E. *et al.* Compensated electron and hole pockets in an underdoped high-$T_c$ superconductor *Phys. Rev. B* **81**, 214524/1-17 (2010).

[48] LeBoeuf, D., Doiron-Leyraud, N., Vignolle, B., Sutherland, M., Ramshaw, B. J., Levallois, J., Daou, R., Laliberté, F., Cyr-Choinière, O., Chang, J., Jo, Y. J., Balicas, L., Liang, R., Bonn, D. A., Hardy, W. N., Proust, C., and Taillefer, L. Lifshitz critical point in the cuprate superconductor $YBa_2Cu_3O_y$ from high-field Hall effect measurements. *Phys. Rev. B* **83**, 054506/1-14 (2011).

[49] Rourke, P. M. C. *et al.* A detailed de Haas−van Alphen effect study of the overdoped cuprate $Tl_2Ba_2CuO_{6+\delta}$ *New J. Phys.* **12**, 105009/1-29 (2010).

[50] Helm, T., Kartsovnik, M. V., Bartkowiak, M., Bittner, N., Lambacher, M., Erb, A., Wosnitza, J., and Gross, R. Evolution of the Fermi Surface of the Electron-Doped High-Temperature Superconductor $Nd_{2-x}Ce_xCuO_4$ Revealed by Shubnikov−de Haas Oscillations *Phys. Rev. Lett.* **103**, 157002/1-4 (2009).

[51] Carrington, A. Quantum oscillation studies of the Fermi surface of iron-pnictide superconductors *Rep. Prog. Phys.* **74**, 124507/1-13 (2011).

[52] Vignolle, B., Carrington, A., Cooper, R. A., French, M. M. J., Mackenzie, A. P., Jaudet, C., Vignolles, D., Proust, C. & Hussey, N. E., Quantum oscillations in an overdoped high-$T_\text{c}$ superconductor. *Nature* **455**, 952-955 (2008).

[53] Vignolle, B., Vignolles, D., LeBoeuf, D., Lepault, S., Ramshaw, B., Liang, R., Bonn, D.A., Hardy, W. N., Doiron-Leyraud, N., Carrington, A., Hussey, N. E., Taillefer, L., Proust, C. Quantum oscillations and the Fermi surface of high-temperature cuprate superconductors. *Comptes Rendus Physique* **12**, 446-460 (2011).

[54] Fazekas, P. Lecture Notes on Electron Correlation and Magnetism, (World Scientific, Singapore 1999).

[55] Pines, D., Noziéres, P. *The Theory of Quantum Liquids: Normal Fermi Liquids* (Redwood City, Addison-Wesley 1989).

[56] Bennett, A. J., Falicov, L. M. $g$ Factor in Metallic Zinc *Phys. Rev.* **136**, A998-A1002 (1964).

[57] Proshin, J. N., Useinov, N. K. The spin flip in the theory of magnetic breakdown: magnetoresistance *Physica B* **173**, 386-388 (1991).

[58] Garcia-Aldea, D. & Charkravarty, S. Multiple quantum oscillation frequencies in YBa$_2$Cu$_3$O$_{6+\delta}$ and bilayer splitting. *New J. Phys.* **12**, 105005 (2010).

[59] Reynoso, A. A., Usaj, G., Balseiro, C. A. Magnetic breakdown of cyclotron orbits in systems with Rashba and Dresselhaus spin-orbit coupling *Phys. Rev. B.* **78**, 115312/1-9 (2008).

[60] Harrison, N. Near Doping-Independent Pocket Area from an Antinodal Fermi Surface Instability in Underdoped High Temperature Superconductors *Phys. Rev. Lett.* **107**, 186408/1-4 (2011).

[61] Harrison, N. & Sebastian, S. E. Fermi surface reconstruction from bilayer charge ordering in the underdoped high temperature superconductor YBa$_2$Cu$_3$O$_{6+x}$. *New J. Phys.* **14**, 095023 (2012).

[62] Andersen, O. K., Liechtenstein, A. I., Jepsen, O., and Paulsen, F. LDA energy bands, low-energy Hamiltonians, $t'$, $t''$, $t_\perp(k)$ and $J_\perp$. *J. Phys. Chem. Solids* **56**, 1573-1591 (1995).

[63] Lee, P. A. From high temperature superconductivity to quantum spin liquid: progress in strong correlation physics. *Rep. Prog. Phys.* **71**, 012501 (2008).

[64] Chakravarty, S., Laughlin, R. B., Morr, D. K., and Nayak, C. Hidden order in the cuprates. *Phys. Rev. B* **63**, 094503 (2001).

[65] Wang, Z-Q., Kotliar, G., & Wang, X.-F. Flux-density wave and superconducting instability of the staggered-flux phase. *Phys. Rev. B* **42**, 8690-8693 (1990).

[66] Emery, V. J. & Kivelson, S. A. Importance of phase fluctuations in superconductors with small superfluid density. *Nature* **374**, 434-437 (1994).

[67] Tranquada, J. M., Sternlieb, B. J., Axe, J. D., Nakamura, Y. & Uchida, S. Evidence for stripe correlations of spins and holes in copper oxide superconductors *Nature* **375**, 561-563 (1995).

[68] Chubukov, A. V. & and Morr, D. K. Electronic structure of underdoped cuprates *Phys. Rep.* **288** 355-387 (1997).

[69] Varma, C. M. Non-Fermi-liquid states and pairing instability of a general model of copper oxide metals. *Phys. Rev. B* **55**, 14554-14580 (1997).

[70] Schmalian, J., Pines, D., & Stojkovich, B. Weak Pseudogap Behavior in the Underdoped Cuprate Superconductors. *Phys. Rev. Lett.* **80**, 3839- (1998).

[71] Kivelson, S. A., Bindloss, I. P., Fradkin, E., Oganesyan, V., Tranquada, J. M., Kapitulnik, A. & Howald, C. How to detect fluctuating stripes in the high-temperature superconductors *Rev. Mod. Phys.* **75**, 120141 (2003).

[72] Choy, T. -P. & Phillips, Ph. Doped Mott Insulators Are Insulators: Hole Localization in the Cuprates *Phys. Rev. Lett.* **95**, 196405 (2005).

[73] Kyung, D. -B. et al. Pseudogap induced by short-range spin correlations in a doped Mott insulator. *Phys. Rev. B* **73**, 165114 (2006).

[74] Li, C., Zhou, S., Wang, Z. Inhomogeneous states with checkerboard order in the $t-J$ model. *Phys. Rev. B* **73**, 060501(R)/1-4 (2006).

[75] Zaanen, J., Chakravarty, S., Senthil, T., Anderson, P., Lee, P. A., Schmalian, J. A., Imada, M., Pines, D., Randeria, M., Varma, C., Vojta, M., and Rice, T. M. Towards a complete theory of high $T_c$ *Nat. Phys.* **2**, 138-143 (2006).

[76] Monthoux, P., Pines, D., & Lonzarich, G. G. Superconductivity without phonons. *Nature* **450**, 1177-1183 (2007).

[77] Tewari, S., Zhang, C., Yakovenko, V. M., & Das Sarma, S. Time-Reversal Symmetry Breaking by a (d+id) Density-Wave State in Underdoped Cuprate Superconductors. *Phys. Rev. Lett.* **100**, 217004 (2008).

[78] Anderson, P. W. & Casey, P. A. Hidden Fermi liquid: self-consistent theory for the normal state of high-$T_c$ superconductors. *Phys. Rev. Lett.* **106**, 097002 (2009).

[79] Berg, E., Fradkin, E., Kivelson, S. A. & Tranquada, J, Striped superconductors: How the cuprates intertwine spin, charge and superconducting orders. *New J. Phys.* **11**, 115004 (2009).

[80] Sedrakyan, T. A. & Chubukov, A. V. Pseudogap in underdoped cuprates and spin-density-wave fluctuations. *Phys. Rev. B* **81**, 174536 (2010).

[81] Sushkov, O. P. Magnetic properties of lightly doped antiferromagnetic $YBa_2Cu_3O_y$ http://arXiv.org/abs/1105.2102v1 (2011).

[82] Rice, T. M., Yang, K.-Y. & Zhang, F. C. A Phenomenological Theory of the Anomalous Pseudogap Phase in Underdoped Cuprates. *Rep. Prog. Phys.* **75**, 016502 (2012).

[83] Scalapino, D. J. A common thread: The pairing interaction for unconventional superconductors. *Rev. Mod. Phys.* **84**, 1383 (2012).

[84] Sachdev, S. & La Placa, R. Bond Order in Two-Dimensional Metals with Antiferromagnetic Exchange Interactions. *Phys. Rev. Lett.* **111**, 027202 (2013).

[85] Wang, Y. & Chubukov, A. Superconductivity at the onset of the spin-density-wave order in a metal. *Phys. Rev. Lett.* **110**, 127001 (2013).

[86] Tremblay, A. -M. S. Strongly correlated superconductivity. Autumn School on Correlated Electrons: Emergent Phenomena in Correlated Matter, Forschungszentrum Jülich, Germany (2013).

[87] Hosur, P., Kapitulnik, A., Kivelson, S. A., Orenstein, J., Raghu, S. Kerr effect as evidence of gyrotropic order in the cuprates. *Phys. Rev. B* **87**, 115116/1-8 (2013).

[88] Gull, E., Parcollet, O., & Millis, A. J. Superconductivity and the Pseudogap in the two-dimensional Hubbard model. *Phys. Rev. Lett.* **110**, 216405 (2013).

[89] Kivelson, S. A., Fradkin, E., and Emery, V. J. Electronic liquid-crystal phases of a doped Mott insulator. *Nature* **393**, 550-553 (1998).

[90] Sau, J.D. & Sachdev, S. Mean field theory of competing orders in metals with antiferromagnetic exchange interactions. *Phys. Rev. B* **89**, 075129 (2014).

[91] Laughlin, R. B. Hartree-Fock computation of the high-$T_c$ cuprate phase diagram. *Phys. Rev. B* **89**, 035134 (2014).

[92] Meier, H., Pepin, C., Einenkel, M., & Efetov, K. B. Cascade of phase transitions in the vicinity of a quantum critical point. *Phys. Rev. B* **89**, 195115 (2014).

[93] Alexandrov, A. S. & Bratkovsky, A. M. de Haas-van alphen effect in canonical and grand canonical multiband fermi liquid *Phys. Rev. Lett.* **76** 1308-1311 (1996).

[94] Carrington, A, and Yelland, E. A. Band-structure calculations of Fermi-surface pockets in ortho-II YBa$_2$Cu$_3$O$_{6.5}$ *Phys. Rev. B* **76**, 140508(R)/1-4 (2007).

[95] Elfimov, I. S., Sawatsky, G. A., and Damascelli, A. Theory of Fermi-surface pockets and correlation effects in underdoped YBa$_2$Cu$_3$O$_{6.5}$. *Phys. Rev. B* **77**, 060504(R)/1-4 (2008).

[96] Melikyan, A & Vafek, O. Quantum oscillations in the mixed state of d-wave superconductors. *Phys. Rev. B* **78**, 020502(R) (2008).

[97] Varma, C. M. Magneto-oscillations in underdoped cuprates. *Phys. Rev. B* **79**, 085110 (2009).

[98] Senthil, T. & Lee, P. A. Synthesis of the phenomenology of the underdoped cuprates. *Phys. Rev. B* **79**, 245116 (2009).

[99] Pereg-Barnea, T., Weber, H., Refael, G. & Franz, M. Quantum oscillations from Fermi arcs. *Nature Phys.* **6**, 44-49 (2010).

[100] Oh, H., Choi, H. J., Louie, S. G. and Cohen. M. L. Fermi surfaces and quantum oscillations in the underdoped high-$T_c$ superconductors YBa$_2$Cu$_3$O$_{6.5}$ and YBa$_2$Cu$_4$O$_8$ *Phys. Rev. B* **84** 014518 (2011).

[101] Ong, N. P. Geometric interpretation of the weak-field Hall conductivity in two-dimensional metals with arbitrary Fermi surface *Phys. Rev. B* **43**, 193-201 (1991).

[102] Swanson, J. A. Saturation Hall Constant of Semiconductors. *Phys. Rev.* **99**, 1799-1807 (1955).

[103] Ziman, J. M. *Principles of the Theory of Solids*, (Cambridge University Press 1979).

[104] Stark, R. W., Reifenberger, R. Quantitative theory for the quantum interference effect in the transverse magnetoresistance of Pure Magnesium. *J. Low Tem. Phys.* **26**, 763-817 (1977).

[105] Harrison, N. *et al.* Magnetic breakdown and quantum interference in the quasi-two-dimensional superconductor $\kappa-(\text{BEDT}-\text{TTF})_2\text{Cu}(\text{NCS})_2$ in high magnetic fields. *J. Phys.-Cond. Matt.* **8**, 5415-5435 (1996).

[106] Basov, D. N. & Timusk, T. Electrodynamics of high-$T_c$ superconductors. *Rev. Mod. Phys.* **77**, 721-779 (2005).

# Extended Data

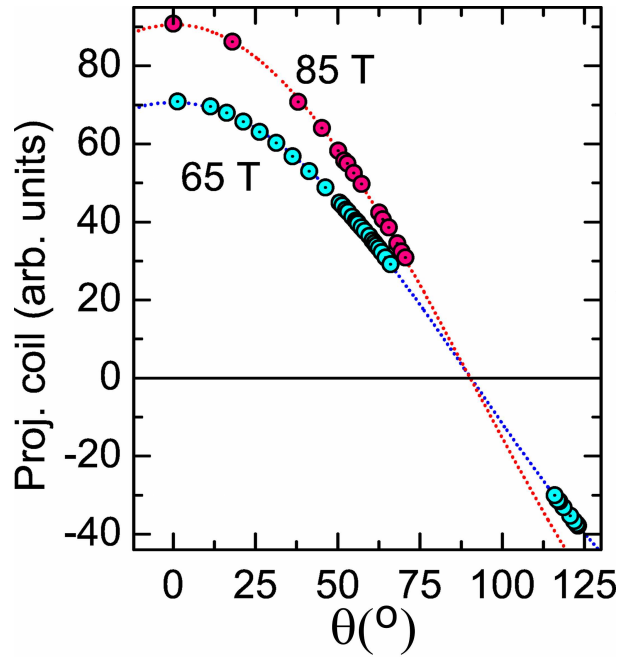

Extended Data Figure 1: **Measured projection of the magnetic field along the crystalline $c$-axis of the sample**. Circles indicate the maximum $B\cos\theta$ measured at 65 T (cyan) and 85 T (red), obtained by means of a projection coil, while the dashed lines represent fits to a cosine function. The angular error is $< 0.2°$ for $\theta \leq 66°$ and $\approx 0.2°$ for $68° \leq \theta \leq 71°$. $\theta = 0°$, 1.3°, 11.3°, 12°, 16.3°, 18°, 21.3°, 26.3°, 31.3°, 36.3°, 38°, 41.3°, 45.2°, 46.3°, 48°, 49°, 49.4°, 50.1°, 50.6°, 51.4°, 51.5°, 52°, 52.3°, 52.5°, 52.9°, 53.1°, 54.4°, 54.9°, 55.5°, 56°, 56.2°, -56.95°, 57.2°, -57.4°, -58.15°, 58.2°, -59.4°, 59.6°, 60.6°, 61.2°, -61.4°, 61.7°, 62.5°, 62.6°, -62.7°, -63.2°, 63.4°, 63.7°, -64.1°, 64.5°, 65.5°, 66°, 66.3°, 68.1°, 69.4° and 70.6°. Negative $\theta$ angles refer to a measured equivalent (180-$|\theta|$) angle as shown.

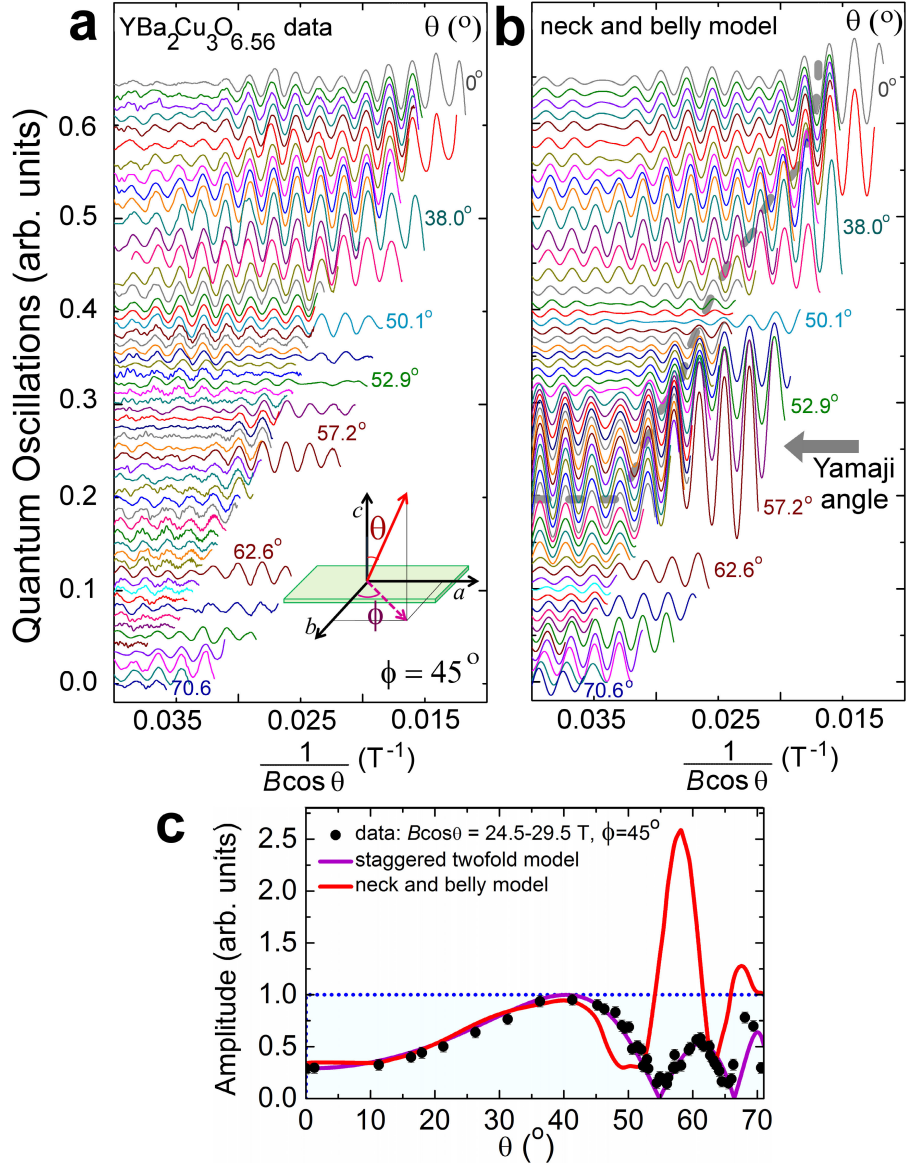

Extended Data Figure 2: **Experimental quantum oscillations for different angles compared with simulations for a neck and belly model**. **a**, Measured oscillations in the contactless resistivity. **b**, Simulated oscillations at the same angles and fields as **a** for two Fermi surface cylinders exhibiting a fundamental neck and belly warping, for parameters used in ref.[36] (listed in Extended Data Table 3) to simulate the restricted experimental range within the dashed line. Data in **a** and simulations in **b** have been scaled by $\exp\frac{100\,\text{T}}{B\cos\theta}$ for visual clarity. **c**, Symbols represent the absolute value of the cross-correlation between the quantum oscillation data in **a** with a simple sinusoid $\exp\left(i\frac{2\pi F}{B\cos\theta}\right)$. $F$ is matched to the periodicity of the oscillations at $\theta = 38°$, where a single frequency dominates the measured quantum oscillations. Coloured lines indicate a simulation $\propto R_{\text{w}} R_{\text{s}}$ for a staggered twofold model using parameters in Extended Data Table 1 (magenta), and a neck and belly model using parameters from ref.[36] in Extended Data Table 3 (red). While the anti-resonance in the vicinity of $\theta = 60°$ yielded by the staggered twofold model is in good agreement with the experimental data, the striking Yamaji resonance in the vicinity of $\theta = 60°$ yielded by the neck and belly model is in marked contrast to experiment.

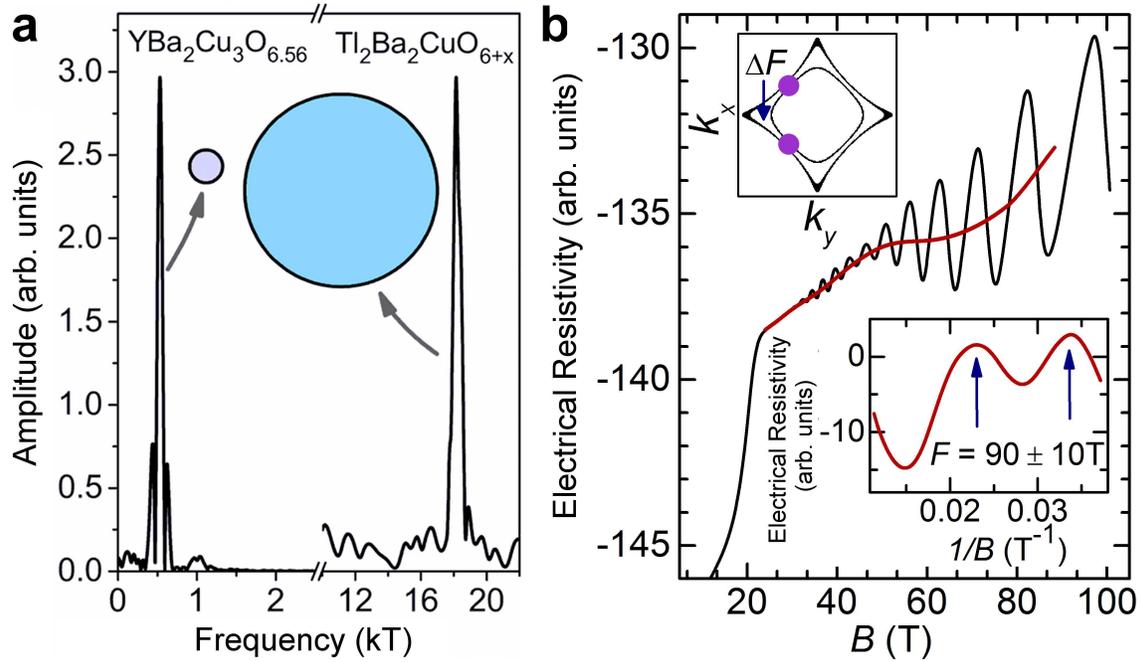

Extended Data Figure 3: **Quantum oscillations extending to 100 T magnetic field**. **a**, Schematic illustration of the small and large Fermi surface pocket sizes (and quantum oscillation Fourier frequencies) in underdoped $YBa_2Cu_3O_{6+x}$ and overdoped $Tl_2Ba_2CuO_{6+\delta}$, respectively.[16,52] **b**, Contactless electrical resistivity of $YBa_2Cu_3O_{6.56}$ measured to 100 T, showing the resistive transition (at $\approx 20$ T) and quantum oscillations. The dominant quantum oscillations of 530 T can be seen to be superimposed on slowly varying oscillations (red line), which we extract in the lower inset by subtracting the dominant oscillations and a linear background. The slowly varying oscillations are consistent with a low frequency of $90 \pm 10$ T. The upper inset shows the bilayer-split pockets expected for charge order in which the difference in area between two magnetic breakdown junctions corresponds to a frequency of $\approx 90$ T.

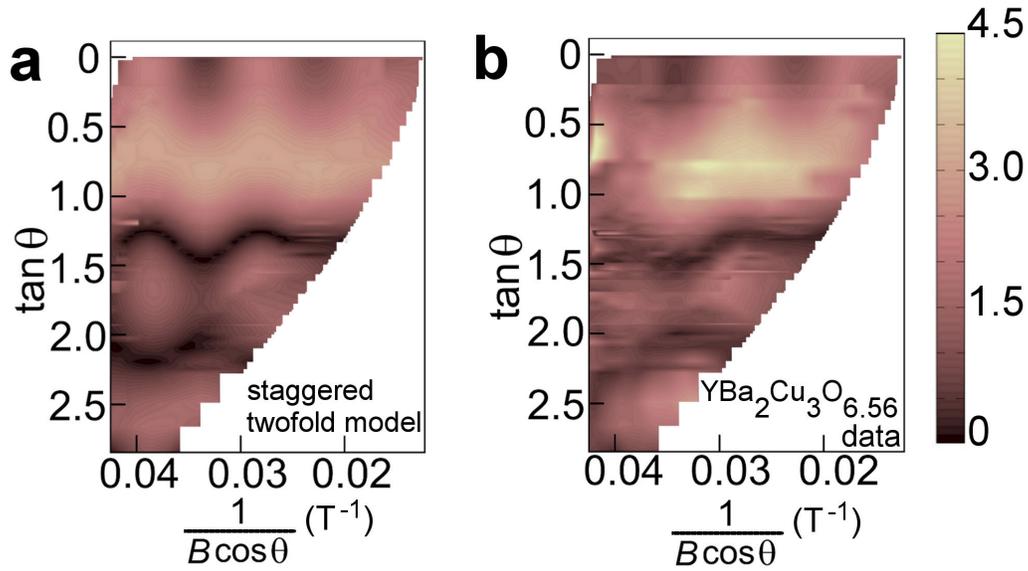

Extended Data Figure 4: **Contour representation of a staggered twofold Fermi surface model compared to the experimental data**. **a**, Contour plot of the simulated quantum oscillation amplitude for a staggered twofold Fermi surface geometry represented by equations 2 and 4, using parameters in Extended Data Table 1 and shown in Fig. 4c. **b**, Contour plot of experimentally measured quantum oscillation amplitude; good agreement is seen with the model in **a**. The quantum oscillation amplitude is indicated by the colour scale (in arbitrary units) in the reciprocal field-angle plane; for clarity the ordinate is given as $\tan\theta$.

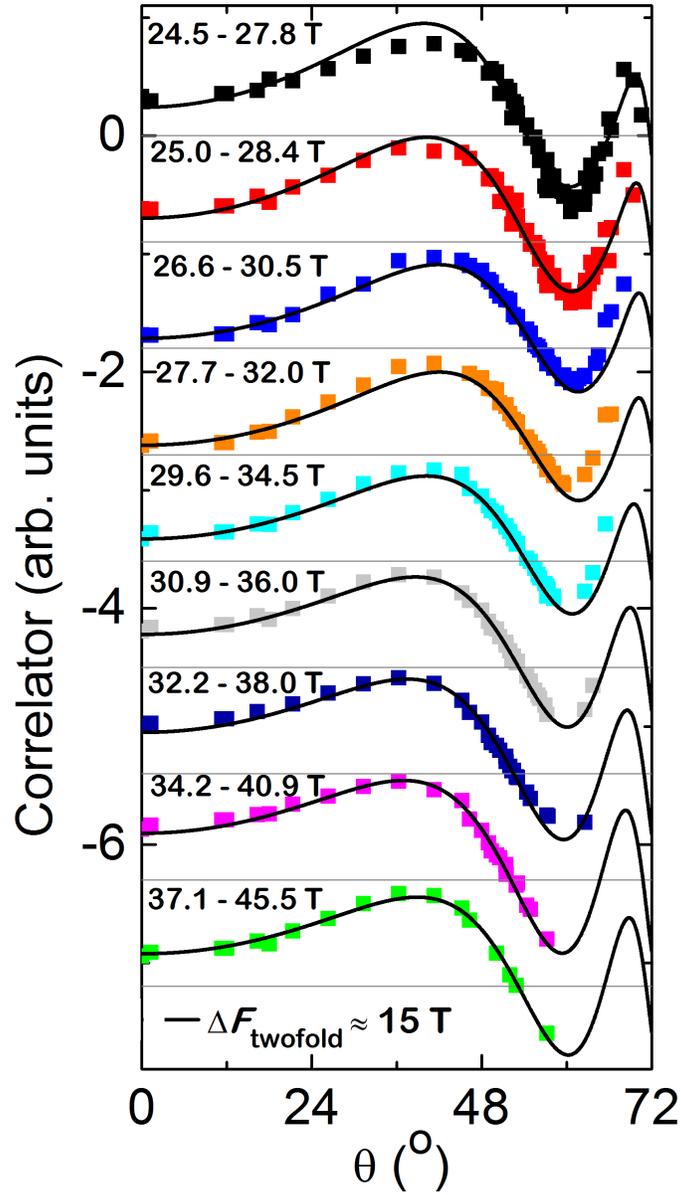

Extended Data Figure 5: **Cross-correlation measured over different field ranges, compared to simulations for a staggered twofold Fermi surface geometry**. Real component of the cross-correlation between the quantum oscillation data over fixed ranges of $B\cos\theta$ for a range of measured $\theta$ angles, with a simple sinusoid $\cos(2\pi F/B\cos\theta + \phi)$. $F$ and $\phi$ are matched to the periodicity and phase of the oscillations at $\theta = 38°$, where a single frequency dominates the measured quantum oscillations. Black lines indicate the simulation (proportional to $R_\mathrm{w}R_\mathrm{s}$), where $\Delta F_\mathrm{twofold} \approx 15$ T is the depth of the modulation, while square symbols indicate the experimental cross-correlation.

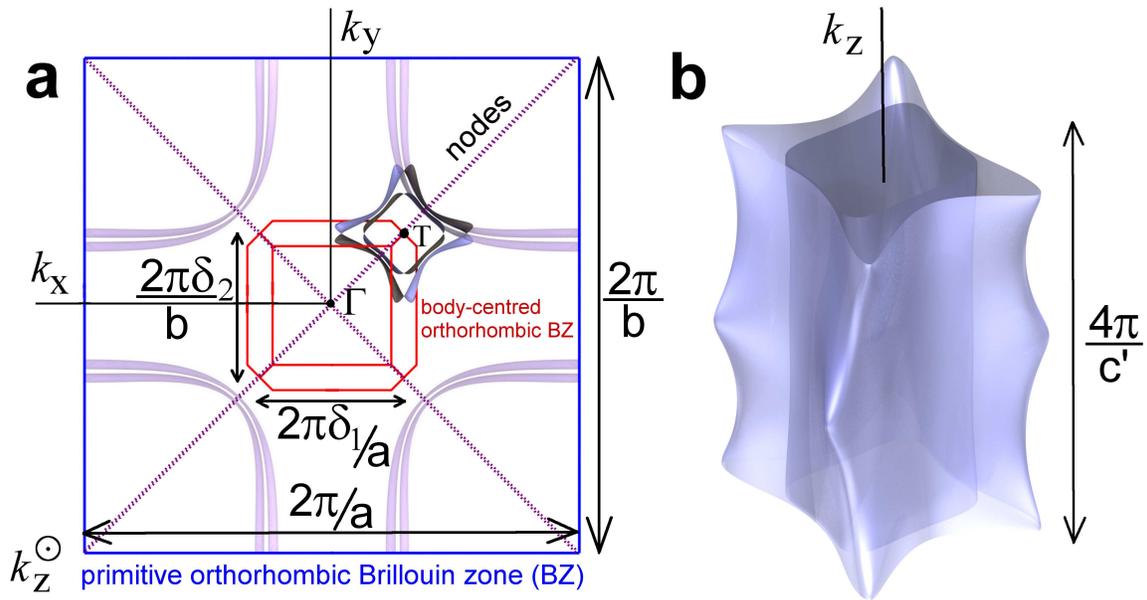

Extended Data Figure 6: **Schematic of a nodal Fermi surface from charge order that staggered perpendicularly to the bilayers.** **a**, Reconstruction of the Brillouin zone, with one instance of the pocket location indicated at the 'T' point in relation to the original Fermi surface (purple) and nodes in the superconducting wave function (from refs.[8,9,60,61]). Here the concentric arrangement of Fermi surfaces arises from bilayer splitting. The in-plane shape of the Fermi pocket shown here is an illustration based on the model in refs.;[8,9,60,61] the corrugation shown is significantly accentuated for clarity. **b**, A three-dimensional view of **a** to illustrate the staggered twofold geometry. While this schematic assumes achirality, a chiral model is not ruled out, as for instance proposed in ref.,[87] where the form of order breaks mirror symmetry within each plane.

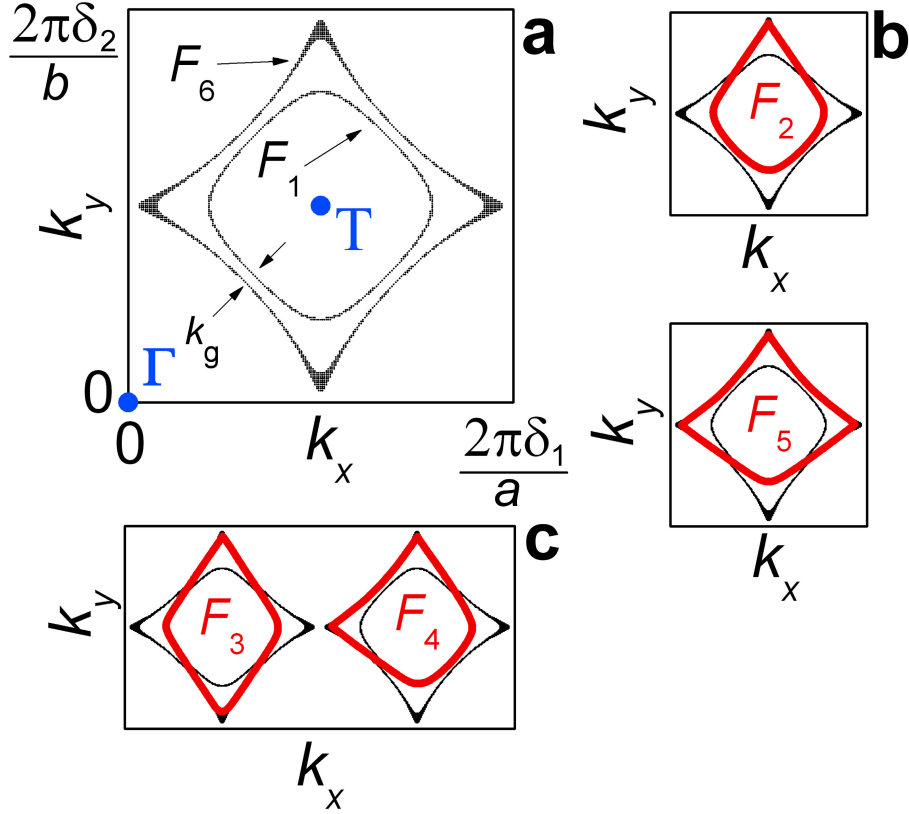

Extended Data Figure 7: **Schematic of the Brillouin zone cross-section, showing magnetic breakdown orbits in a charge ordering scheme.** A cut through the $k_z = 0$ plane of the Brillouin zone shows the six possible orbits resulting from magnetic breakdown tunnelling in a bilayer charge ordering scheme,[16,61] an illustrative Fermi pocket shape similar to Extended Data Fig. 6 is shown. **a** The two Fermi surface cross-sections of frequency $F_1 = F_0 - 2\Delta F_{\text{split}}$ and $F_6 = F_0 + 2\Delta F_{\text{split}}$ that can result from bilayer splitting with in-plane ordering wave vectors $\mathbf{Q}_1^{\parallel} = 2\pi(\pm\frac{\delta_1}{a}, 0)$ and $\mathbf{Q}_2^{\parallel} = 2\pi(0, \pm\frac{\delta_2}{b})$. The $\Gamma$ and T symmetry points of the body-centered orthorhombic Brillouin zone of the charge order superstructure are depicted in blue. The gap separating bonding and antibonding surfaces is expected to be smallest at the nodes.[62] Panels **b**, **c**, **d** and **e** show the range of possible magnetic breakdown orbits, $F_2 = F_0 - \Delta F_{\text{split}}$, $F_3 = F_0$, $F_4 = F_0$ and $F_5 = F_0 + \Delta F_{\text{split}}$, as listed in Table 2.

| parameter | description | value |
|---|---|---|
| $F_0$ | quantum oscillation frequency | 534 T |
| $\Delta F_{\text{twofold}}$ | staggered twofold warping frequency | 15 T |
| $\Delta F_{\text{split}}$ | bilayer splitting frequency | 90 T |
| $m_\parallel^*$ | quasiparticle effective mass | 1.6 $m_{\text{e}}$ (fixed) |
| $B_0$ | magnetic breakdown field | 2.7 T |
| $\Lambda$ | Dingle damping parameter | 132 T |
| $a_0$ | absolute amplitude | 0.131 MHz |
| $g_{\parallel\square}^*$ | g-factor 1 | 2.12 |
| $g_{\parallel\diamond}^*$ | g-factor 2 | 0.35 |
| $\xi_\square$ | g-factor anisotropy 1 | 1.43 |
| $\xi_\diamond$ | g-factor anisotropy 2 | 0.15 |

Extended Data Table 1: **Model parameters for a staggered twofold Fermi surface model.** Tabulated values used in equations 2 and 4 to simulate the quantum oscillation waveform, yielding excellent agreement with experiment as a function of $B$, $\theta$ and $\phi$ (Fig. 4 and Extended Data Figs. 2**c** and 4). The effective mass is taken to be a fixed quantity, having been determined independently from temperature-dependent measurements.[47] The parameters are the same for all the orbits, except for those denoted by subscripts $\square$ and $\diamond$, which each correspond to a subset of orbits as defined in the text. Because of the multiple frequencies in the model, it is not possible to uniquely identify the g-factor; the values of $g_j^*$ and $\xi_j^*$ here represent parameters used for the simulation.

| orbit $F_j$ | $N_j$ | $R_{\mathrm{MB}}$ |
|---|---|---|
| $F_1 = F_0 - 2\Delta F_{\mathrm{split}}$ | 1 | $(1-P)^2$ |
| $F_2 = F_0 - \Delta F_{\mathrm{split}}$ | 4 | $-P(1-P)$ |
| $F_3 = F_0$ | 2 | $P^2$ |
| $F_4 = F_0$ | 4 | $-P(1-P)$ |
| $F_5 = F_0 + \Delta F_{\mathrm{split}}$ | 4 | $-P(1-P)$ |
| $F_6 = F_0 + 2\Delta F_{\mathrm{split}}$ | 1 | $(1-P)^2$ |

Extended Data Table 2: **Magnetic breakdown amplitude damping for the orbits shown in Extended Data Fig. 7**. The magnetic breakdown network corresponds to a split Fermi surface geometry from charge order, as shown in Extended Data Fig. 7. A high magnetic breakdown tunneling probability ($P$) causes the amplitude of the $F_1$ and $F_6$ orbits to become very weak, as seen in Fig. 3**a** (ref.[16]).

| parameter | description | value |
|---|---|---|
| $F_1$ | quantum oscillation frequency 1 | 478 T |
| $F_2$ | quantum oscillation frequency 2 | 526 T |
| $\Delta F_{\mathrm{neck-belly},1}$ | Fermi surface corrugation frequency 1 | 37.7 T |
| $\Delta F_{\mathrm{neck-belly},2}$ | Fermi surface corrugation frequency 2 | 3.5 T |
| $\gamma_1$ | quantum oscillation phase 1 | 3.5 |
| $\gamma_2$ | quantum oscillation phase 2 | 1.1 |
| $m_1^*$ | quasiparticle effective mass 1 | 1.5 $m_{\mathrm{e}}$ |
| $m_2^*$ | quasiparticle effective mass 1 | 1.7 $m_{\mathrm{e}}$ |
| $\Lambda_1$ | Dingle damping parameter 1 | 127.8 T |
| $\Lambda_2$ | Dingle damping parameter 2 | 159.8 T |
| $a_1$ | absolute amplitude 1 | 0.238 MHz |
| $a_2$ | absolute amplitude 2 | 0.339 MHz |
| $g_{\parallel 1}^*$ | $g$-factor 1 | 1.40 |
| $g_{\parallel 2}^*$ | $g$-factor 2 | 1.88 |

Extended Data Table 3: **Model parameters for a fundamental neck and belly Fermi surface model**. Tabulated values from ref.[36] used in equations 1 and 3 to simulate the waveform of the oscillations in Fig. 2**c** and Extended Data Fig. 2**b** and **c** for two cylinders with isotropic $g$-factors. The absolute amplitudes $a_1$ and $a_2$ listed here have been adjusted to match the amplitude in the present experiment.